\documentclass{emulateapj}

\usepackage{multirow}
\usepackage{graphics}
\usepackage{amsmath, amsthm, amssymb}

\newcommand{\sech}{{\rm sech}}

\begin{document}

\title{Star Formation in Disk Galaxies. I. Formation and Evolution of Giant Molecular Clouds via Gravitational Instability and Cloud Collisions}

\author{Elizabeth J. Tasker}
\author{Jonathan C. Tan}
\affiliation{Department of Astronomy, University of Florida, Gainesville, FL 32611, USA}

\begin{abstract}
We investigate the formation and evolution of giant molecular clouds
(GMCs) in a Milky-Way-like disk galaxy with a flat rotation curve. We
perform a series of 3D adaptive mesh refinement (AMR) numerical
simulations that follow both the global evolution on scales of $\sim
20$~kpc and resolve down to scales $\lesssim 10$~pc with a multiphase
atomic interstellar medium (ISM). In this first study, we omit star
formation and feedback, and focus on the processes of gravitational
instability and cloud collisions and interactions. We define clouds as
regions with $n_{\rm H}\geq 100\:{\rm cm^{-3}}$ and track the
evolution of individual clouds as they orbit through the galaxy from
their birth to their eventual destruction via merger or via
destructive collision with another cloud. After $\sim 140$~Myr a large
fraction of the gas in the disk has fragmented into clouds with masses
$\sim 10^6\:M_\odot$ and a mass spectrum similar to that of Galactic GMCs.
The disk settles into a quasi steady state in which
gravitational scattering of clouds keeps the disk near the threshold
of global gravitational instability. The cloud collision time is found
to be a small fraction, $\sim 1/5$, of the orbital time, and this is
an efficient mechanism to inject turbulence into the clouds. This
helps to keep clouds only moderately gravitationally bound, with
virial parameters of order unity. Many other observed GMC properties,
such as mass surface density, angular momentum, velocity dispersion,
and vertical distribution, can be accounted for in this simple model
with no stellar feedback.
\end{abstract}


\maketitle
\section{Introduction}

The formation of stars in disk galaxies is one of the most basic
processes controlling galactic evolution. While there are many other
important effects, such as galactic interactions and infall of diffuse
gas, ultimately a large fraction of gas settles in rotationally
supported disks, where the majority of the stellar population is
born. The appearance of many galaxies is dominated by the light from
massive, short-lived stars whose spatial distribution is controlled by
this star formation process. The injection of heavy elements from
winds and supernovae will occur mostly from this disk environment.

Empirical correlations have been found between the star formation rate
(SFR), gas content and global galactic dynamics. Based on a sample of
about 100 nearby galaxies and circumnuclear starburst disks, Kennicutt
(1998) found relatively simple relations between the globally-averaged
star formation rate per unit area, $\bar{\Sigma}_{\rm sfr}$, and the
total (\ion{H}{1} and $\rm H_2$) gas mass surface density,
$\bar{\Sigma}_{\rm gas}$. They can be related via
\begin{equation}
\label{sfr1}
\bar{\Sigma}_{\rm sfr} = A_{\rm sfr} \bar{\Sigma}_g^{\alpha_{\rm sfr}},
\end{equation}
with $A_{\rm sfr}=(2.5\pm0.7)\times 10^{-4}\:M_\odot\:{\rm yr^{-1}\:kpc^{-2}}$
and $\alpha_{\rm sfr} = 1.4\pm0.15$. Alternatively, an equally good fit
to the data is given by
\begin{equation}
\label{sfr2}
\bar{\Sigma}_{\rm sfr}\simeq B_{\rm sfr} \bar{\Sigma}_g \Omega_{\rm out},
\end{equation}
where $B_{\rm sfr}=0.017$ and $\Omega_{\rm out}$ is the orbital
angular frequency at the outer radius that is used to perform the disk
averages. This last relation implies that a fixed fraction, about
10\%, of the gas is turned into stars every outer orbital timescale of
the star-forming disk.

Martin \& Kennicutt (2001) argued that the outer edge of the
star-forming disk is set by the boundary between the gravitationally
unstable $Q\la 1$ inner disk and the gravitationally stable $Q\ga1$
outer region, where $Q$ is the Toomre (1964) stability parameter
\begin{equation}
\label{Q}
Q \equiv \frac{\kappa \sigma_g}{\pi G \Sigma_g},
\end{equation}
where $\sigma_g$ is the 1D gas velocity
dispersion in the disk plane and $\kappa$ is the epicyclic frequency:
\begin{equation}
\label{kappa}
\kappa=\sqrt{2}\frac{v_c}{r}\left(1+\frac{r}{v_c}\frac{dv_c}{dr}\right)^{1/2}=\sqrt{2}\frac{v_c}{r}(1+\beta)^{1/2}.
\end{equation}
Here, $v_c$ is the circular velocity at a particular galactocentric
radius $r$, and $\beta\equiv d\:{\rm ln}v_c / d\:{\rm ln}r$, which is
$0$ for a flat rotation curve. The precise value of $Q$ below which
gas becomes unstable is also affected by the destabilizing influence
of the potential due to a stellar disk (Jog \& Solomon 1984, Jog 1996,
Kim \& Ostriker 2007). Schaye (2004) argued that the edges of star
forming disks may rather be set by the location where much of a
galaxy's atomic interstellar medium transitions from the cold ($\sim
300$~K) to the warm ($T\simeq 8000$~K) phase.

When galaxies are examined on smaller ($\sim$kpc) scales, e.g. by
taking azimuthal averages, star formation relations similar to
eqs. (\ref{sfr1}) \& (\ref{sfr2}) are found (Wong \& Blitz 2002,
Kennicutt et al. 2007, Leroy et al. 2008, Bigiel et al. 2008),
although there is still debate as to whether it is the total or just
the molecular gas mass that is most fundamental for controlling the
star formation rate (Blitz \& Rosolowsky 2006, Leroy et al. 2008,
Bigiel et al. 2008).

\subsection{GMCs: Formation, Evolution, Observed Properties}\label{S:GMCs}

As we move to smaller scales ($\sim100$~pc) we see, based mostly on
surveys of CO emission in the Milky Way (Solomon et al. 1987; Dame,
Hartmann \& Thaddeus 2001; Jackson et al. 2006) that most of the
dense, cold, potentially star-forming gas is organized into giant
molecular clouds (GMCs) with masses $\sim10^6\:M_\odot$ and radii
$\sim30$~pc (with these particular values of mass and size
corresponding to volume-averaged number densities of H nuclei of
$n_{\rm H}=260\:{\rm cm^{-3}}$ for a spherical cloud). Given that most
of the Galactic molecular gas is organized in GMCs and that GMCs have
approximately equal mass atomic envelopes (Blitz 1990), the results of
Wolfire et al. (2003) for the radial distribution of Galactic
molecular and atomic gas indicate that a large fraction, $\sim 1/3$,
of the total gas mass in the Milky Way inside the solar circle is
associated with these structures. Many questions concerning GMCs are
still debated, including whether they are long or short-lived compared
to their free fall timescale,
\begin{equation}\label{eq:tff}
t_{\rm ff} = \left(\frac{3\pi}{32G\rho}\right)^{1/2} = 4.35 \times 10^6 \left(\frac{n_{\rm H}}{100\:{\rm cm^{-3}}}\right)^{-1/2}{\rm yr},
\end{equation}
and whether they are gravitationally bound (McKee \& Ostriker 2007,
and references therein). 

Observations of GMCs are most complete in the Milky Way interior to
the solar circle (Solomon et al. 1987; Williams \& McKee 1997; Heyer
et al. 2008). Except for the vicinity of the Galactic center, there is
a dearth of clouds in the inner $\sim2$~kpc. GMCs occupy a very thin
vertical distribution in the Galaxy: Bronfman et al. (2000) derived a
half-intensity height of $z_{1/2}\simeq 60$~pc; Stark \& Lee (2005)
derived a vertical scale height of $\lesssim 35$~pc. 

Stark \& Brand (1989) derived a 1D cloud to cloud RMS velocity
dispersion of $7.8\pm0.5\:{\rm km\:s^{-1}}$ from a study of GMCs
within 3~kpc of the Sun, although this estimate includes small scale
streaming motions.

Based on the $^{12}$CO survey data of Solomon et al. (1987), Williams
\& McKee (1997) derived a cloud mass function of the form
\begin{equation}
\label{cloud_mf}
\frac{d{\cal N}_c}{d {\rm ln} M_c} = {\cal N}_{\rm cu} \left( \frac{M_c}{M_u} \right)^{-\alpha_{c}}
\end{equation}
for $M_c\leq M_u$, with $d {\cal N}_c(M_c)$ being the number of clouds
with masses in the range $[M_c,M_c(1+d{\rm ln}M_c)]$. Based on clouds
in the mass range of a few $\times 10^5\:M_\odot$ to a few $\times
10^6\:M_\odot$, Williams \& McKee (1997) estimated that the population
of GMCs inside the solar circle is described by $\alpha_c=0.6$, ${\cal
N}_{\rm cu}=63$ and $M_u=6\times 10^6\:M_\odot$. If the observational
surveys have higher degrees of incompleteness at lower masses,
Williams \& McKee estimated $\alpha_c=0.85$, ${\cal N}_{\rm cu}=25$ and
$M_u=6\times 10^6\:M_\odot$. Note these mass estimates did not include
any atomic gas that might be associated with the GMCs, which could be
a substantial fraction ($\sim 1/2$) of the total mass bound to the
cloud (Blitz 1990).

The mean mass surface density, $\Sigma_c$, of Galactic GMCs was derived
by Solomon et al. (1987) to have a median value of about
$200\:M_\odot\:{\rm pc}^{-2}$. Recently, Heyer et al. (2008) derived a
median value of $42\:M_\odot\:{\rm pc^{-2}}$, based on an LTE analysis
of $^{13}$CO survey data. They estimate the true values are larger
because of subthermal excitation and abundance variations and are in
the range $80-120\:M_\odot\:{\rm pc^{-2}}$. Again these estimates do
not include any atomic gas associated with the clouds.

Bertoldi \& McKee (1992) define the virial parameter, 
\begin{equation}\label{alpha_vir}
\alpha_{\rm vir} \equiv \frac{5 \sigma_{c}^2 R_{c,A}}{G M_c},
\end{equation}
where $\sigma_{c}$ is the mass-averaged 1D velocity dispersion of
the cloud, which includes thermal and nonthermal contributions via
$\sigma_{c}\equiv (c_s^2+\sigma_{\rm nt,c}^2)^{1/2}$ where
$\sigma_{\rm nt,c}$ is the 1D RMS velocity dispersion about the cloud's
center of mass velocity, and $R_{c,A}$ is a measure of cloud radius
based on its projected area. The virial parameter is a measure of the
ratio of the kinetic to gravitational energy of a cloud. For a
uniform, spherical cloud, $\alpha_{\rm vir}=1$ implies the total
kinetic energy of the cloud is half the magnitude of the gravitational
energy. Nonspherical and nonuniform density distributions typically
make only modest effects, parameterized via the dimensionless factor
$a$ in the gravitational energy equation $W=-(3/5)a GM_c^2/R_{c,A}$:
Bertoldi \& McKee (1992) estimate that cloud major to minor axis
ratios of a factor of 3 only cause $a$ to deviate from unity by
$\lesssim 10\%$, while a power law density profile $\rho\propto
r^{-3/2}$ results in $a=1.25$. Observationally, Heyer et al. (2008)
find a median value of $\alpha_{\rm vir}$ of about unity.


Phillips (1999) considered the rotational properties of Galactic GMCs,
finding a significant spread in the directions of the angular momentum
vectors (as derived from radial velocity gradients and as projected on
the plane of the sky), indicating that a substantial fraction of GMCs
rotate in a retrograde direction with respect to Galactic rotation.

Observations of GMCs in other galaxies tend to find they have similar
properties as Galactic GMCs, including their velocity dispersions and
virial parameters (Bolatto et al. 2008). Rosolowsky et al. (2003)
measured the rotational properties of large GMCs in M33, finding
relatively small values of specific angular momentum, and a
substantial population ($\sim 40\%$) of clouds with rotation
directions retrograde with respect to that of the galaxy. Fukui et
al. (2008) studied 164 GMCs in the Large Magellanic Cloud, concluding
these clouds were also close to virial equilibrium.

The star formation activity within GMCs appears highly clustered.
Most stars are born from star-forming clumps that turn into star
clusters (Lada \& Lada 2003), with initial radii $\sim1$~pc. Here the
local overall star formation efficiency is relatively high,
$\epsilon_*\equiv M_*/M_{\rm gas} \sim 0.1 - 0.5$, where $M_*$ is the
total mass of stars formed from the mass of gas $M_{\rm gas}$ that
occupies the same volume as forming star cluster. Conversely, most of
the volume of GMCs is not actively forming stars, perhaps because of
magnetic field support (Crutcher 2005).

Theoretical work on the formation of GMCs has led to different schools
of thought (McKee \& Ostriker 2007, and references therein).
`Top-down' formation mechanisms suggest GMCs form via large-scale
gravitational or magnetic disk instabilities (e.g. Kim, Ostriker, \&
Stone 2003; Shetty \& Ostriker 2006; Glover \& Mac Low 2007), whereas
`bottom-up' processes have GMCs forming via agglomeration from
inelastic collisions of clouds (e.g. Kwan 1979) or from turbulent or
colliding flows (e.g. Bergin et al. 2004; Heitsch et al. 2008). It is
possible that both processes may be important depending on the
galactic environment (Dobbs 2008). If GMCs are relatively long-lived
and contain a large fraction of the total ISM, then the formation and
destruction of GMCs to and from the atomic phase may be less important
than the interactions between those already existing (Tan 2000): in
other words, the nature of a particular cloud may change more
drastically via merging collisions with other GMCs than via flows into
or out of the atomic phase.

In this paper, as a first step towards understanding galactic star
formation rates, we will examine the formation and evolution of GMCs
in the context of the global dynamics of a flat rotational curve disk galaxy.



\subsection{Theories of Disk Galaxy Star Formation Rates}

Numerous theories have been proposed to explain the observed
galaxy-scale, Kennicutt star formation relations. One group of
theories is based on the growth rate of gravitational perturbations in
a disk. The timescale for perturbation growth can be expressed as
$t_{\rm grow}\propto (G \rho_g)^{-0.5}$ (e.g. Larson 1988;
Elmegreen 1994; Wang \& Silk 1994), and so $\rho_{\rm sfr}
\propto\rho_g/t_{\rm grow}\propto\rho_{g}^{1.5}$. Assuming a
constant disk scale height, we obtain equation (\ref{sfr1}) with
$\alpha_{\rm sfr}=1.5$ for local mass surface densities. However,
disk-averaged quantities will depend on the radial gas
distribution. We can also express $t_{\rm grow} \sim \sigma_g/(\pi
G \Sigma_g) \sim Q/\kappa$. Perturbation growth via swing
amplification in a differentially rotating disk occurs at a similar
rate (e.g. Larson 1988). By assuming star formation self-regulates
and keeps $Q$ constant, Larson (1988) and Wang \& Silk (1994) predicted
$\Sigma_{\rm sfr}\propto \Sigma_{g}/t_{\rm grow} \propto\Sigma_{g}
\Omega$, since $\kappa \propto \Omega$, for disks with flat rotation
curves. Li, Mac Low \& Klessen (2006) presented isothermal smooth
particle hydrodynamic (SPH) simulations of disk galaxies, from which
they concluded rate of the nonlinear development of gravitational instability
determines the local and global Kennicutt relations.

However, these theories that involve the star formation rate being set
by the growth rate of large scale gravitational instabilities leading
to GMC formation, have difficulty in explaining why a large mass
fraction of the gas is already organized into GMCs. If the GMCs are
forming rapidly in $\sim1$ dynamical timescale the mass flux into (and
out of) GMCs would be $\sim 100$ times greater than that from GMCs
into star clusters (Zuckerman \& Evans 1974). To understand global
star formation rates, it seems more reasonable to look for processes
that create the star-forming parsec-scale clumps within GMCs, rather
than the processes that create the GMCs themselves.

The spatial correlation of star formation with large scale spiral
structure in some disk galaxies motivates theories for the triggering
of star formation during the passage of gas through density
waves. Wyse (1986) and Wyse \& Silk (1989) propose a SFR law of the
form
\begin{equation}
\label{wyse}
\Sigma_{\rm sfr}\propto \Sigma_{g}^{\alpha_{\rm sfr}}(\Omega-\Omega_{p})
\end{equation}
where $\Omega_{p}$ is the pattern frequency of the spiral density
wave. In the limit of small $\Omega_{p}$ and for $\alpha_{\rm sfr}=1$
we recover equation (\ref{sfr2}). 
Increased cloud collision rates and increased perturbation growth
rates in the arms, where $Q$ is locally lowered, have been suggested
as the star formation triggering mechanism (e.g. Dobbs 2008). Kim,
Kim, \& Ostriker (2008) examined the development of thermal
instabilities in galactic spiral shocks.

One prediction of these theories is a correlation of SFR with the
density wave amplitude. However this is not observed (Elmegreen \&
Elmegreen 1986; McCall \& Schmidt 1986, Kennicutt 1989). Furthermore
such theories have difficulty explaining star formation in galaxies
where there is a lack of organized star formation features, as in
flocculent spirals (e.g. Thornley \& Mundy 1997a, 1997b; Grosbol \&
Patsis 1998). GMCs are present and SFRs are similar to those systems
where star formation is organized into spiral patterns. This suggests
stellar disk instabilities, which create spiral density waves, and gas
instabilities, which lead to GMCs and large-scale star formation, are
decoupled (Kennicutt 1989; Seiden \& Schulman 1990).

Tan (2000) proposed a model in which the majority of star formation in
disk galaxies occurs in the pressurized regions triggered by
GMC-GMC collisions. GMCs are observed to occupy a very thin vertical
distribution in the Galaxy (Stark \& Lee 2005),
which is similar to the actual sizes of the clouds. In this
effectively two dimensional system, collisions are set by galactic
shear at impact parameters of about the tidal radius of the clouds
(Gammie, Ostriker, \& Jog 1991), and for a $Q\sim1$ disk with a
relatively large mass fraction, $\sim 1/2$, in GMCs and associated
gas, collisions are expected to occur approximately every 20\% of an
orbital time. In this way, for flat rotation curve galaxies, equation
(\ref{sfr2}) is recovered, and the cloud collision mechanism is the
link between the global galactic dynamics and the parsec-scale
star-forming clumps of GMCs. One prediction of this model is a
reduction in the star formation rate per orbital time for galaxies
with lower rates of shear, i.e. those with rising rotation curves that
are closer to solid body rotation. This model also requires GMCs to be
relatively long-lived: i.e. several tens of Myr.

Krumholz \& McKee (2005) proposed that the star formation rate is
regulated by turbulence in a galaxy's molecular gas, so that a fixed
fraction of the turbulent, molecular gas is converted to stars per
free fall time of the GMC. To recover equation (\ref{sfr2}), one has
to assume that GMC dynamical timescales are a fixed fraction of
galactic dynamical timescales and that GMC mass fractions are constant
in different galactic systems (see also Wada \& Norman 2007).

\subsection{Numerical Studies of GMC Formation and Disk SFRs}

The large range of spatial scales involved from GMCs to global
galactic properties (a range of $\sim 10^4$ from $\lesssim$ 10
parsec-scale clouds to the galactic scale) has made it extremely
difficult for simulations to encompass GMC formation on global
scales. Work that does perform this is either limited to two
dimensions (Shetty \& Ostriker 2008) or has to assume a fixed
two-phase medium for the ISM (Dobbs 2008). While the clouds do largely
reside in the plane of the disk (which is in essence a two-dimensional
system), cloud collisions and feedback can eject gas from the surface,
a process that controls the pressure of the ISM from which the clouds
are forming (McKee \& Ostriker 1977; Cox 2005). Similarly, a fixed
phase ISM results in clouds being created from gas with a
predetermined structure, from which is it harder to discern the main
physical effects controlling their formation and
evolution. Three-dimensional models that contained a self-consistent
multiphase atomic ISM on global scales were performed by Tasker \&
Bryan (2006, 2008) and Wada \& Norman (2007), but either did not reach
the resolution needed to probe below the most massive GMCs (e.g. the
simulations of Tasker \& Bryan (2008) had a limiting resolution of
25-50\,pc), or only considered the inner galaxy (e.g. the simulations
of Wada \& Norman (2007) were restricted to $r <
2.56$\,kpc). Robertson \& Kravtsov (2008) presented a global SPH
simulation, with similar effective resolution to the simulation of
Tasker \& Bryan (2008) (Kravtsov, private comm.) of star formation in
disk galaxies including a multiphase ISM, and subgrid models for
various forms of stellar feedback and molecular hydrogen formation and
destruction (see also Gnedin, Tassis, \& Kravtsov 2008).

Local models can resolve down to smaller scales and, if set up in a
shearing box, can approximate the effects of galactic shear on cloud
formation and evolution (e.g. Kim \& Ostriker 2001, 2006, 2007). Other
examples of local models have focussed on cloud formation from imposed
colliding flows (e.g. Heitsch et al. 2008), including the effects of
magnetic fields (Heitsch, Stone, \& Hartmann 2009). Other groups have
studied cloud formation in the context of the local interplay between
supernova feedback and a turbulent, multiphase ISM (e.g. Slyz et
al. 2005). Simulations including the nonequilibrium formation and
destruction of $\rm H_2$ molecules have been carried out by Glover \&
Mac Low (2007a,b).

While these local models are very important for studying many aspects
of GMC formation and evolution, they cannot explore the global
evolution of GMCs as they orbit through the disk or measure how GMC
(and star formation) properties are related to the more
readily-observed global galactic properties such as mean gas mass
surface density. Trends with galactocentric radius are also easier to
study in global models.

In this paper we present results from a three-dimensional global
galaxy ($\sim 32$\,kpc box containing a gravitationally unstable
galactic disk with diameter 20~kpc) simulation that tracks the
formation and evolution of clouds in a self-consistent multiphase
atomic ISM. Our main simulation has a limiting resolution of $7.8$~pc.
As discussed below, we do not at this stage include the physics of
star formation and feedback. We also do not include magnetic fields,
cooling below 300~K, or any treatment of the formation and destruction
of molecules. Our definition of ``GMCs'', which we also refer to
interchangeably as ``clouds'', is discussed in detail below, but
basically involves a density threshold of $n_{\rm H}\geq 100\:{\rm
cm^{-3}}$. In real galaxies, the structures that correspond to this
definition will typically be composed of a mixture of atomic and
molecular gas, but we argue that the bulk dynamical properties of the
clouds can still be reasonably well captured by our simulations.


\section{Numerical Techniques}

\subsection{The Code}

The simulations performed in this paper were run using {\it Enzo}; a
three-dimensional adaptive mesh refinement (AMR) hydrodynamics code
(Bryan \& Norman 1997; Bryan 1999; O'Shea et al. 2004). {\it Enzo} has
been used previously in galactic disk simulations (e.g. Tasker \&
Bryan 2006, 2008), where it successfully produced a self-consistent
multiphase atomic interstellar medium (ISM), consisting of a wide
range of densities and temperatures. Grid codes are particularly adept
at modeling multiphase gases since the grid cells form natural
boundaries allowing the gas to evolve with a range of temperatures,
densities and pressures. Particle-based techniques also struggle to
resolve fluid instabilities (Tasker et al. 2008) and can suffer from
over-mixing problems unless specific algorithmic steps are
taken. 

{\it Enzo} evolved the gas using a 3D version of the {\it Zeus}
hydrodynamics algorithm (Stone \& Norman 1992). This scheme uses an
artificial viscosity term to model shocks and the variable associated
with this, the quadratic artificial viscosity, was set to 2.0 (the
default) for all simulations. Radiative gas cooling followed the solar
metallicity cooling curve of (Sarazin \& White 1987) down to
temperatures of $T = 10^4$\,K and extends down to $T = 300$\,K using
rates from (Rosen \& Bregman 1995). These temperatures take us to the
upper end of the atomic cold neutral medium (Wolfire et al. 2003). In
this study we do not include the formation and destruction of
molecules or any cooling processes below 300~K. By a combination of
dust and molecular cooling, gas in real GMCs reaches temperatures of
$\sim 10$\,K, more than an order of magnitude below our minimum
radiative cooling temperature. However, since we are not able to
resolve the detailed internal structure of clouds and their internal
turbulence (a typical cloud with a diameter of $\sim 100$~pc would
only have $\sim 13$ cells across each linear dimension in our highest
resolution simulation) and since we are not including magnetic
pressure support, we view our temperature floor of 300~K as being
equivalent to imposing a minimum 1D signal speed equal to the sound
speed $c_s = (\gamma P/\rho)^{1/2} = (\gamma k T/ \mu)^{1/2} = 1.80
(T/300{\rm K})^{1/2} \rm km\:s^{-1}$, where $\gamma=5/3$, $\mu=1.273
m_p$ (for an assumed $n_{\rm He}=0.1n_{\rm H}$). This signal speed is
somewhat smaller than observed internal velocity dispersions of
GMCs. In fact we will see that, even though our simulation does not
include feedback, our simulated GMCs typically attain internal
nonthermal (i.e. turbulent) velocity dispersions much greater than
this minimum.

The galaxy was modeled in a 3D simulation box of side $32$\,kpc with
isolated gravitational boundary conditions and outflow fluid
boundaries. For our main, high-resolution simulation, the root grid
was $256^3$ with an additional four levels of refinement, producing a
minimum cell size of $7.8$\,pc. To examine how simulation results
depend on resolution, we also carried out medium and low-resolution
runs with a minimum cell sizes of 15.6~pc and 31.2~pc, respectively.
Refinement of a cell occurred when the Jeans length drops below four
cell widths, in accordance with the criteria suggested by (Truelove et
al. 1997) for resolving gravitational instabilities. Resolution of the
fragmentation is discussed further in
Section~\ref{sec:gravcollapse}. The disk was allowed to evolve for
$324$\,Myr, about 1.3 orbital periods at $r = 8.0$\,kpc (see
Section~\ref{sec:initial_disk}) and substantially longer than the
local free fall time
of the initial mid-plane gas at this location, which was about 60~Myr.

In this paper, the first of a series, we examine the formation and
evolution of GMCs in a flat rotation curve galactic disk without the
presence of star formation and feedback mechanisms, which we defer to
future papers. Since the amount of mass removed by star formation is
expected to be only a few percent per local free fall time (Zuckerman
\& Evans 1974; Krumholz \& Tan 2007), the gas depletion over the course
of the simulation time would be relatively modest. The structure of
the ISM and the GMC population we derive can be regarded as that
resulting from the limiting case when there is very little coupling of
stellar feedback (including FUV heating) to the ISM. Gravitational
scattering of bound clouds in a shearing disk and dissipative cloud
collisions are the processes that will act in our simulations to
extract orbital energy and regulate the ISM structure.

\subsection{The Initial Structure of the Galactic Disk}
\label{sec:initial_disk}


To mimic the present-day Milky Way, the simulated galaxy was set up
as an isolated disk of gas orbiting in a static background potential
which represented both dark matter and a stellar disk component. This
minimized effects from the evolution of the galaxy on the interstellar
structure, allowing us to investigate the formation and evolution of GMCs in a
steady-state environment.  The background potential was chosen to
produce a flat rotation curve at $r \gg r_c$, where the core radius was set to be
$r_c = 0.5$\,kpc. The form of the potential is (Binney \& Tremaine 1987):
\begin{equation}
\Phi = \frac{1}{2}v_{c,0}^2\ln\left[\frac{1}{r_c^2}\left(r_c^2 + r^2 + \frac{z^2}{q_\phi^2}\right)\right],
\end{equation}
where $v_{c,0}$ is the constant circular velocity in the limit of
large radii, here set equal to $200\:{\rm km\:s^{-1}}$, $r$ and $z$
are the radial and vertical coordinates respectively, and the axial
ratio of the potential field is $q_\phi = 0.7$. The form of the
circular velocity of the disk, $v_c$, is then given by:
\begin{equation}
v_c=\frac{v_{c,0}r}{\sqrt{r_c^2 + r^2}}.
\end{equation}

The initial radial profile of the gas mass surface density in the disk
was chosen to give a constant value of the Toomre $Q$ parameter for
gravitational instability (Toomre 1964):
\begin{equation}
\Sigma_g = \frac{\kappa \sigma_g}{\pi G Q},
\end{equation}
where $\sigma_g$, the 1D velocity dispersion of the gas, is equivalent
to the sound speed $c_s$ for the case of an razor thin disk with only
thermal pressure. The particular choices of $Q$ at different radii are discussed below.
For our simulated gas disk we define $\sigma_g \equiv (\sigma_{\rm
nt}^2 + c_s^2)^{1/2}$, averaged over the mass in particular regions of
the disk, where $\sigma_{\rm nt}$ is the 1D velocity dispersion of the
gas motions in the plane of the disk after subtraction of the circular
velocity, i.e. representing nonthermal motions. Our disks are
initialized with $\sigma_{\rm nt}=0$.

The initial vertical profile of the gas was set proportional to
$\sech^2 (z/z_h)$, where $z_h$ is the vertical scale height, which was
assumed to vary, i.e. increase, with galactocentric radius based on
observations of the HI in the Milky Way presented in (Binney \&
Merrifield 1998). At the solar radius of $8$\,kpc, $z_h = 290$\,pc.
Then we have $\Sigma_g = \int^\infty_{-\infty} \rho_0 {\rm
sech}^2(z/z_h) dz = 2 \rho_0 z_h$, where $\rho_0$ is the midplane
($z=0$) density, so that the gas distribution becomes:
\begin{equation}
\rho(r,z) = \frac{\kappa \sigma_g}{2 \pi G Q z_h} \sech^2{\left(\frac{z}{z_h}\right)}.
\label{eq:rho}
\end{equation}
%
%
%

The initial disk profile is divided radially into three sections. In
our main region of interest, between radii of $r = 2 - 10$\,kpc (that
is, encompassing the part of the Galaxy inside the solar circle with
radius of 8~kpc) $\Sigma_g$ is set so that $Q=1$ if $\sigma_g=c_s$ were equal
to 6~km~s$^{-1}$, similar to the observed velocity dispersion of the ISM
(Kennicutt 1998; Stark \& Brand 1989). The actual initial velocity
dispersion (i.e. sound speed) is set
to $c_s = 9.0$\,km~s$^{-1}$, so that $T_{\rm init} \sim 7450$\,K and $Q = 1.5$.
The threshold of gravitational instability is crossed by allowing the
gas to cool. The other regions of the galaxy, from 0 to 2~kpc and from
10 to 12~kpc, are initialized in a similar way, but are designed to be
gravitationally stable with $Q=20$ for a flat rotation curve if $c_s$
were to equal 6~km~s$^{-1}$ (although note the rotation curve is not flat in
the center). Their initial temperatures are set to the same value as
in the main disk region. Even after cooling to the temperature floor
of 300~K, for which $c_s=1.8$\,km~s$^{-1}$, these regions remain
gravitationally stable, except for a small amount of gas that collects
at the galaxy center. Beyond 12~kpc, the disk is surrounded by a
static, very low density medium, that has negligible influence.  Only
the main disk region from 2 to 10~kpc is analyzed in this paper, and
in fact, due to boundary effects, we typically restrict analysis to
radii from 2.5 to 8.5~kpc.

In total, the gas mass was $6 \times 10^9$\,M$_\odot$. Note, this is a
factor of ten smaller than the disks presented in Tasker \& Bryan
(2008).

\subsection{Defining and Analyzing Giant Molecular Clouds}

\begin{figure} 
\begin{center}
\includegraphics[width=\columnwidth]{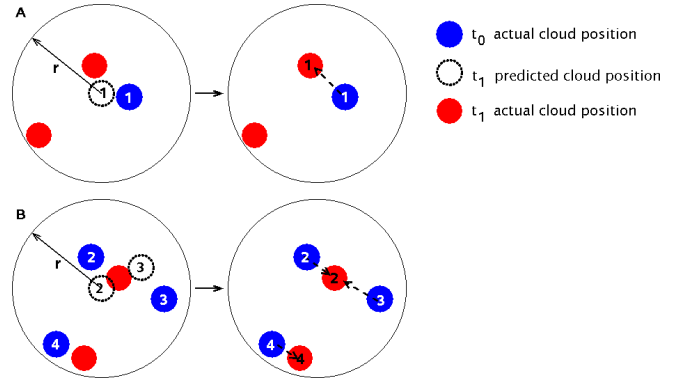} 
\caption{Diagram to illustrate how clouds are tracked from one output
timestep to the next, typically 1~Myr later. Panel (A) shows a simple
example: the predicted position (center of mass) of cloud 1 at time
$t_1$ is shown by the black dotted circle. A volume of radius 50~pc is
searched for clouds present at $t_1$ and the nearest cloud is tagged
as being cloud 1 at time $t_1$. Panel (B) shows a more complex
scenario where two clouds present at time $t_0$ merge to form a single
cloud at time $t_1$. (diagram is illustrative and not to scale).
\label{fig:cloud_tracking}}
\end{center} 
\end{figure}

\begin{figure*} 
\begin{center}
\includegraphics[width=7cm]{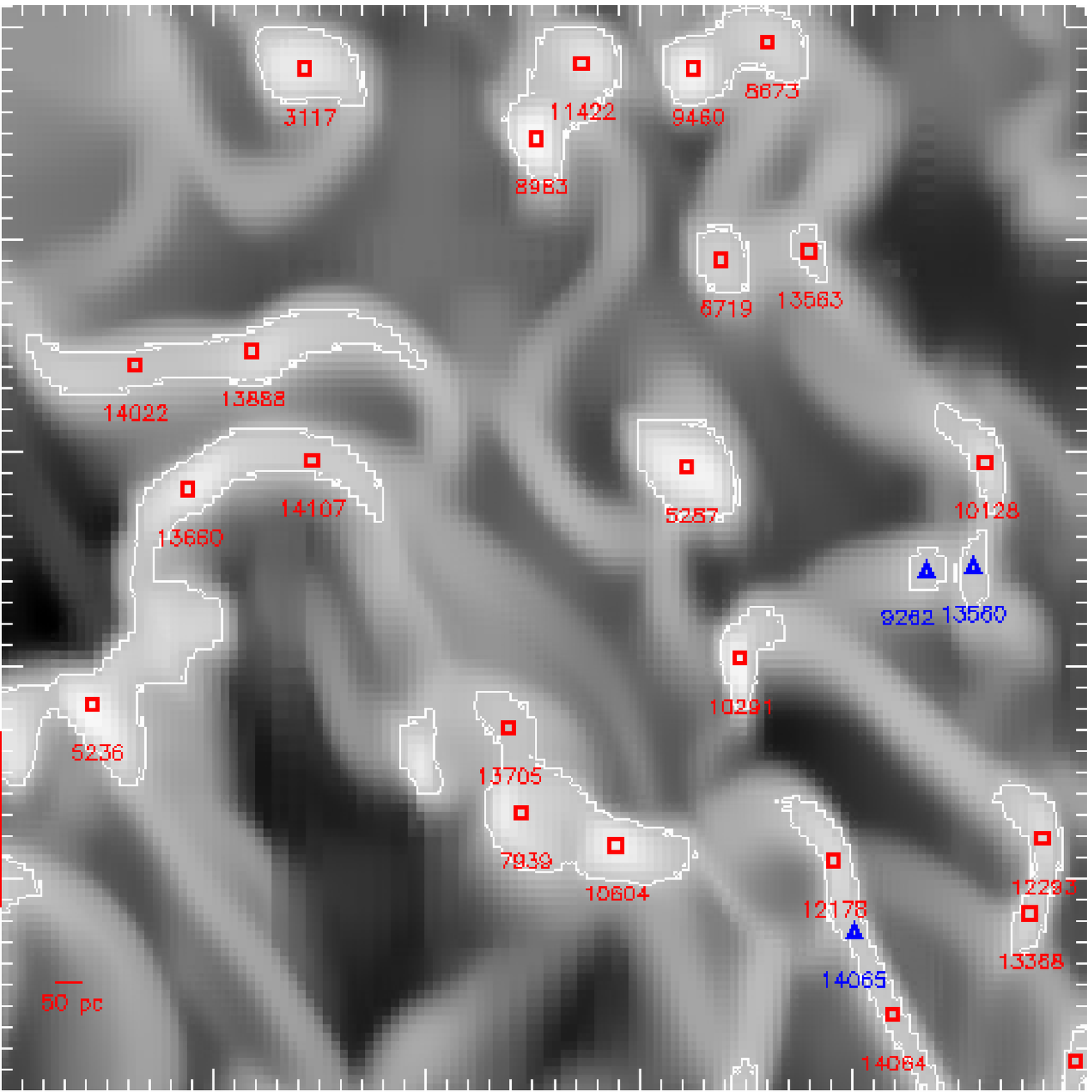}
\includegraphics[width=7cm]{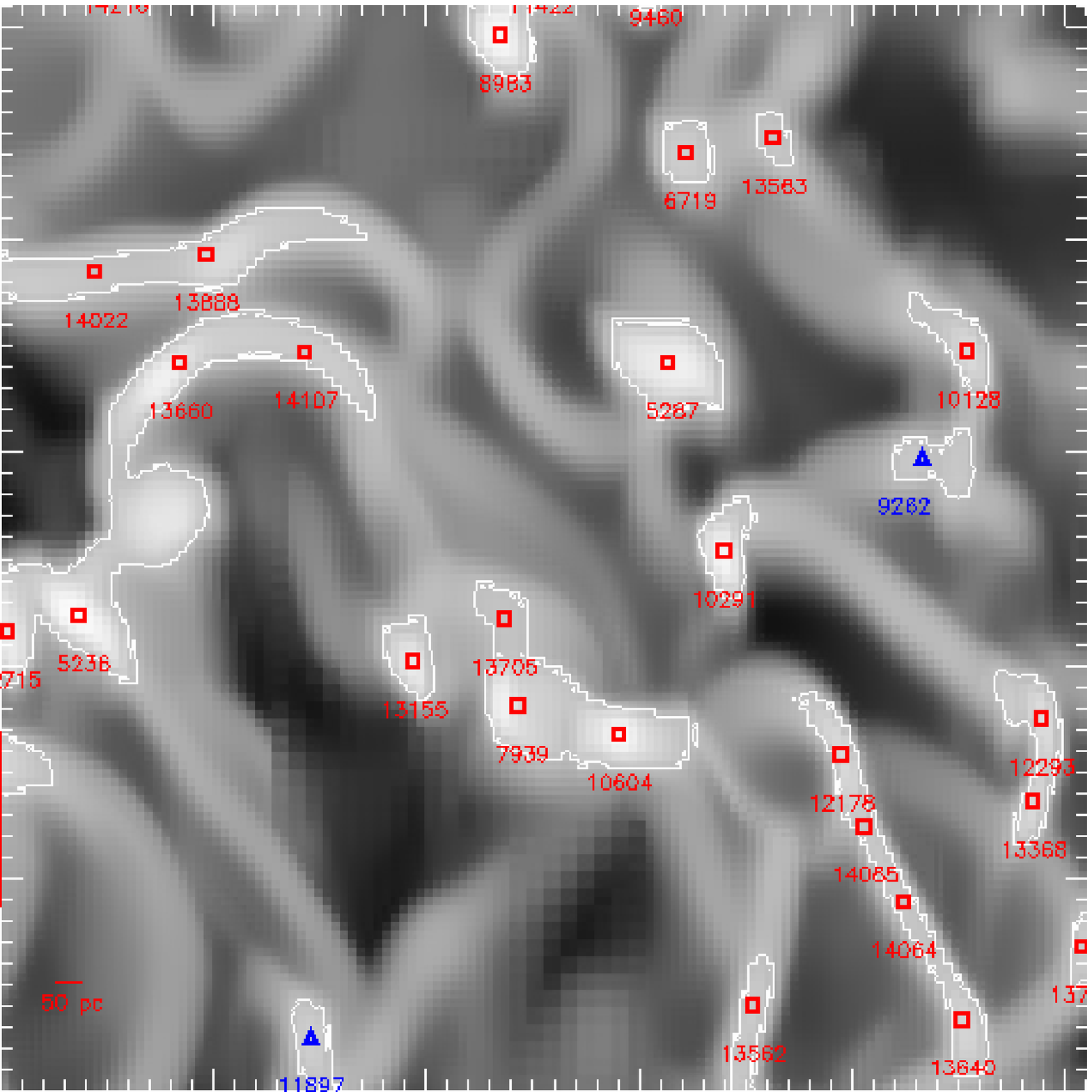}
\caption{Example of cloud tracking in the medium resolution (15.6~pc)
simulation. The left panel shows a 2~kpc square patch of the galactic
plane, 5~kpc from the galactic center, at $t=160$~Myr, 15.6~pc thick
in the z-direction, with the grey scale showing column density through
this slice, i.e. proportional to $n_{\rm H}$ in the cells. Contours
show the cloud definition threshold density of $n_{\rm H,c} =
100$\,cm$^{-3}$, while the numbered squares and triangles show the
center-of-mass position of the clouds located inside the volume of the
slice with prograde and retrograde rotation, respectively. The right
panel shows the same patch at $t=161$~Myr. Galactic orbital motion has
carried clouds and the ISM structure upwards and to the left. The
cloud tracking algorithm successfully tracks clouds as they move, and
records a merger of clouds 9282 and 13550 on the right side of the
images. Note the center of mass of cloud 13155 enters the slice from
above during this time interval, just lower-left of center of the image.
\label{fig:cloud_closeup}}
\end{center} 
\end{figure*}

The clouds in the galactic disk were located using a number density of
H nuclei threshold of $n_{\rm H,c} = 100$\,cm$^{-3}$, similar to the
mean (volume-averaged) densities of typical Galactic GMCs. Also recall
that the formation of molecules is not being followed in this
simulation. When we refer to `clouds' or `GMCs' we are
describing the gas that has achieved densities of $n_{\rm H}\geq
n_{\rm H,c}$ in the atomic phase.  We expect that most of the gas
above this density would form molecules via surface reactions on dust
grains, although there could still be substantial ($\sim$ equal mass)
atomic components present, as are observed in and around Galactic GMCs
(Wannier et al. 1991; Blitz 1990).

The process of locating and tracking the clouds in the disk at a given
time in the simulation is outlined below:

\begin{enumerate}

\item Clouds were identified as peaks in the baryon density field with $n_{\rm H} \geq n_{\rm H,c} = 100$\,cm$^{-3}$. If two peaks were $\leq 4$ minimum cell widths apart, i.e. about 30~pc in our high resolution run, then only the higher density peak was retained for cloud definition.

\item Non-peak cells with $n_{\rm H} \geq n_{\rm H,c}$ were assigned to the nearest peak that was connected to them by cells with $n_{\rm H} \geq n_{\rm H,c}$, i.e. clouds are continuous structures.


\end{enumerate}

Once the cloud had been defined, properties including the cloud's
center-of-mass, velocity and angular momentum were calculated. Multiple clouds
could exist in the same continuous density structure if it contained
more than one well-separated peak. 

To follow the evolution of the clouds, simulation outputs were
analyzed every 1\,Myr and the clouds mapped between outputs with a tag
number assigned to each cloud to follow its life through the
simulation. To map a cloud between times $t_0$ and $t_1$, the code
performed the following steps:

\begin{enumerate}

\item Assuming linear motion, a predicted position of each cloud's
center-of-mass found at time $t_0$ is calculated for the cloud at
$t_1$.

\item A volume of radius 50~pc, larger than the expected deviations
from linear motion due to typical accelerations (e.g. over 1~Myr these
are about 20~pc at $r=2$~kpc due accelerations in the galactic
potential), centered on the predicted position of each cloud is
searched for clouds present at time $t_1$. If multiple clouds are
found in this region at $t_1$, the nearest one is chosen to be
associated with the cloud from $t_0$. In cases where two or more $t_0$
clouds are associated with the same $t_1$ cloud, the nearest one is
matched and the volume around the predicted positions of the other
$t_0$ clouds is searched for alternative candidates.

\item If no clouds at $t_1$ are associated with a $t_0$ cloud, then a
volume with radius equal to $3 \times$ the average radius of the $t_0$
cloud is searched. This allows for large, extended clouds whose radius
may be $\gtrsim 50$\,pc and whose centers may have shifted due to
external perturbations.

\item Any $t_0$ clouds remaining unassigned may have merged with
neighboring $t_0$ clouds. A volume of radius $2 \times$ the average
radius of each $t_1$ cloud is searched for unassigned $t_0$
clouds. This value was chosen to be lower than for the previous step
since a recently merged cloud is likely to be fairly extended. If
found, a merger is declared between the $t_0$ cloud previously
associated with the $t_1$ cloud and the unassigned $t_0$ cloud. The $t_1$
cloud inherits the tag number from the more massive of the two merged
clouds. If there were multiple $t_1$ clouds close to an unassigned
$t_0$ cloud, the closest was chosen. Multiple mergers involving more
than two clouds were possible, though were typically rare.

\item Any $t_0$ clouds remaining unassigned after these steps are
declared to have been destroyed by non-merger processes.

\end{enumerate}

Note that this method assumes that any cloud that has been destroyed
in close proximity to another cloud has suffered a merger. This
assumption works well in the current simulation, which has no stellar
feedback, but it remains to be tested once feedback processes are
operating.

Figure~\ref{fig:cloud_tracking} illustrates two possible examples of
the cloud mapping. The blue circles show clouds at $t = t_0$ and the
red circles show clouds located 1\,Myr later at $t = t_1$. In panel
(A), the 
predicted position of cloud 1 at $t=t_1$ is marked by the black dotted
open circle. The surrounding region with radius 50\,pc is searched for clouds
present at $t_1$ and two are found. The closest of these is identified to
be cloud 1 at $t_1$ whereas the other cloud is either a newly formed
cloud or a different cloud present at $t_0$. Panel (B) shows how the
algorithm deals with a more complex situation involving a
merger. Here, a cloud at $t_1$ is associated with both $t_0$ clouds 2
and 3. The code checks there are no unassigned clouds at $t_1$ that
are a viable match for cloud 2 or 3 but finding none, it declares this
a merger event.  Note, that the cloud detection algorithm does not
allow distinct clouds to be found within four minimum cell widths of
each other.

Figure~\ref{fig:cloud_closeup} shows how the cloud identification and
tracking work in the simulation. The two images show the gas density
in a one-cell thick slice of a square patch of the galaxy disk
midplane 2~kpc across and 5~kpc from the galactic center, taken 1\,Myr
apart. Contour lines mark the density threshold of $n_{\rm H,c} \geq
100$\,cm$^{-3}$ and the squares and triangles (prograde and retrograde
rotators, see \S\ref{sec:cloud_simtime}) show the center-of-mass of
each of the clouds identified, with their tag number written
below. Areas where contours or a peak appear to be visible without a
cloud are due to the cloud's center-of-mass being above or below the
slice shown. The cloud tracking algorithm correctly maps clouds
between the two outputs and shows a merger of clouds 9282 and 13580 on
the right-hand side.

\subsection{Resolving Gravitational Collapse}
\label{sec:gravcollapse}

\begin{figure} 
\begin{center} 
\includegraphics[width=\columnwidth]{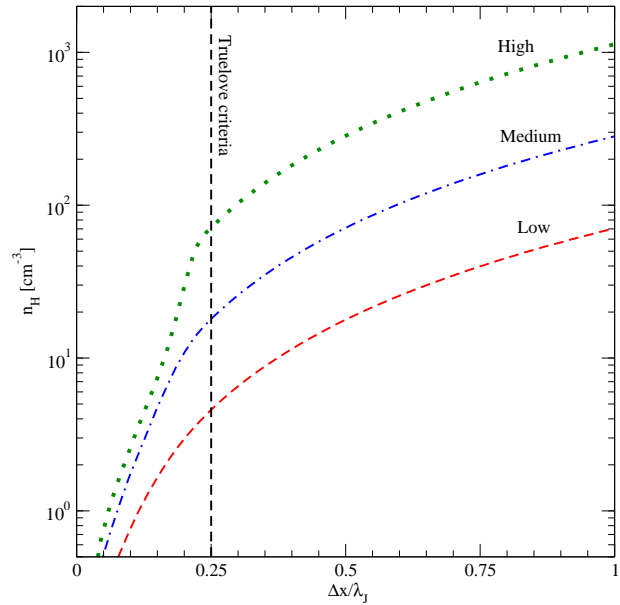}
\caption{Mean density of cells as a function of their Jeans number,
  $N_J=\Delta x / \lambda_J$ for low, medium and high resolution runs
  at $t=250$~Myr. The results at earlier times are similar. Note most
  of the gas with $n_{\rm H}>10\:{\rm cm^{-3}}$ has cooled to the
  temperature floor of 300~K, so for fixed cell size $\Delta x$, we
  have $N_J \propto n_{\rm H}^{1/2}$. This graph shows that in the
  high resolution run fragmentation of structures forming out of
  material with densities up to $n_{\rm H}\simeq 70\:{\rm cm^{-3}}$ is
  well-resolved, and this applies to most clouds forming from ambient
  galactic gas. The fragmentation of the clouds (which have typical
  volume averaged densities of a few hundred) themselves is less
  well-resolved. The effect of resolution on the cloud mass function
  and other properties is examined in \S\ref{sec:cloud_simtime}.}
\label{fig:jeansnumber_rho}
\end{center} 
\end{figure}

Finite computational resources impose a limit on the number of levels
of refinement we are able to employ in the simulation. Cells at the
highest level of refinement are themselves not refined further, which
means they can grow in mass and density to the point that the Truelove
et al. (1997) condition to avoid artificial fragmentation is no longer
satisfied, i.e. that the Jeans length, $\lambda_J$, associated with
gas in each cell should be at least 4 times larger than the size of
the cell, $\Delta x$. In Figure~\ref{fig:jeansnumber_rho} we show the
mean density of cells as a function of their Jeans number, $N_J=\Delta
x / \lambda_J$. This figure indicates that for the high resolution
simulation the Truelove et al. criterion is satisfied for densities up
to $n_{\rm H}\simeq 70\:{\rm cm^{-3}}$, which is close to the
threshold density used to define our GMCs. Thus we expect that the
gravitational fragmentation process that leads to the formation of
GMCs is reasonably well resolved in our simulation. It is not the goal
of this paper to follow the fragmentation inside GMCs, and for this
reason a minimum effective sound speed of about 1.8~km~s$^{-1}$ is
imposed (see above). It is possible that large GMCs may suffer
spurious fragmentation in the simulation, however it should be noted
that these GMCs typically have internal velocity dispersions that are
much larger than the sound speed (see \S\ref{sec:cloud_simtime}). We
examine the effect of simulation resolution on the GMC mass function
and other properties in \S\ref{sec:cloud_simtime}.


%

\section{Global ISM Structure and Evolution}


\begin{figure*}
\begin{center} 
\includegraphics[width=\textwidth]{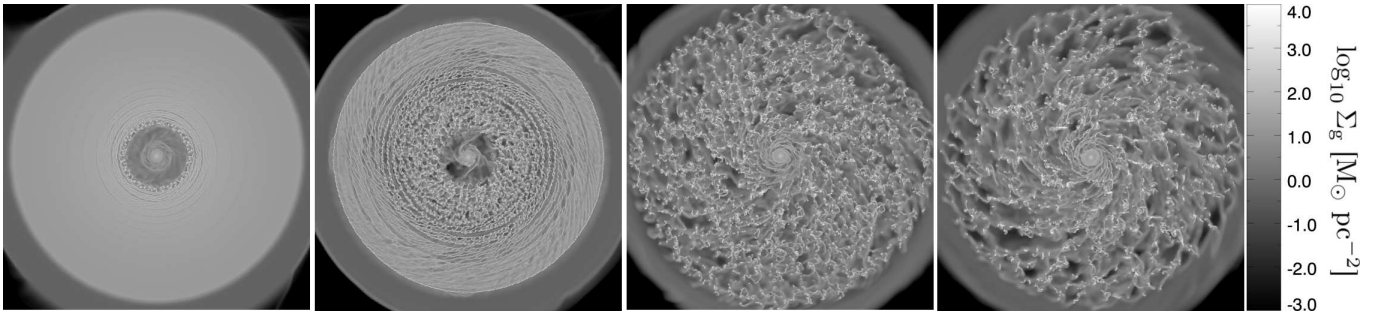}
\caption{Evolution of the galactic disk. Images are 20 kpc across and
show the disk gas mass surface density, $\Sigma_g$ (integrated
vertically over $|z| \leq 1$~kpc) at $t = 50, 100, 200$ and
$300$\,Myr. The formation of rings via the Toomre instability is
evident at earlier times. These rings fragment into individual clouds,
which then suffer interactions via galactic differential rotation. The
properties of the clouds in this fully fragmented stage ($t\gtrsim
140$~Myr) are the focus of this paper.
\label{fig:disk_evol_images}}
\end{center} 
\end{figure*}

\begin{figure*} 
\begin{center} 
\includegraphics[angle = 270, width=\textwidth]{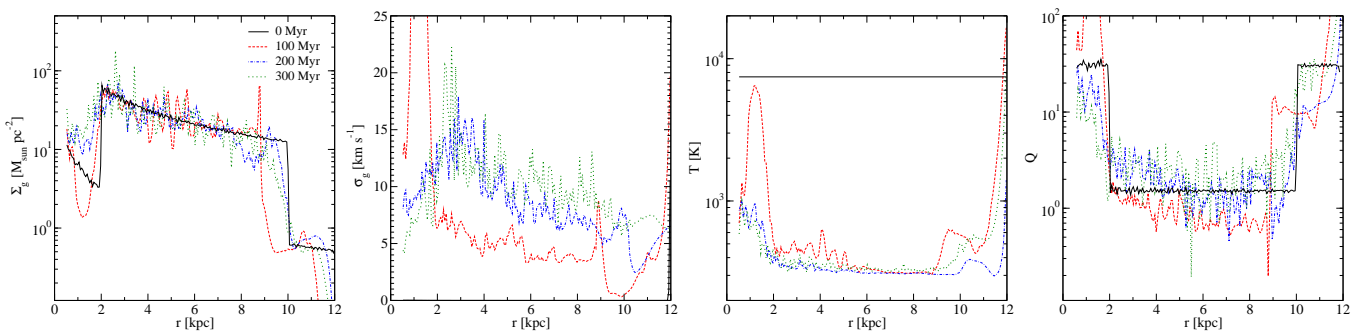}
\caption{Galactic disk azimuthally-averaged (60~pc wide annuli) radial
profiles and their evolution; from left to right: (a) gas mass surface
density, $\Sigma_g= \int_{-1{\rm kpc}}^{+1{\rm kpc}}\rho(z) d z$, (b)
1D gas velocity dispersion, $\sigma_g$, (mass-weighted average over
$-1~{\rm kpc}<z<1~{\rm kpc}$ utilizing only disk plane velocity
components), (c) gas temperature, $T$, (mass-weighted average over
$-1~{\rm kpc}<z<1~{\rm kpc}$), (d) Toomre $Q$ parameter, evaluated
using $\Sigma_g$ and $\sigma_g$.
\label{fig:disk_evol_plots}}
\end{center} 
\end{figure*}

\begin{figure} 
\begin{center} 
\includegraphics[width=\columnwidth]{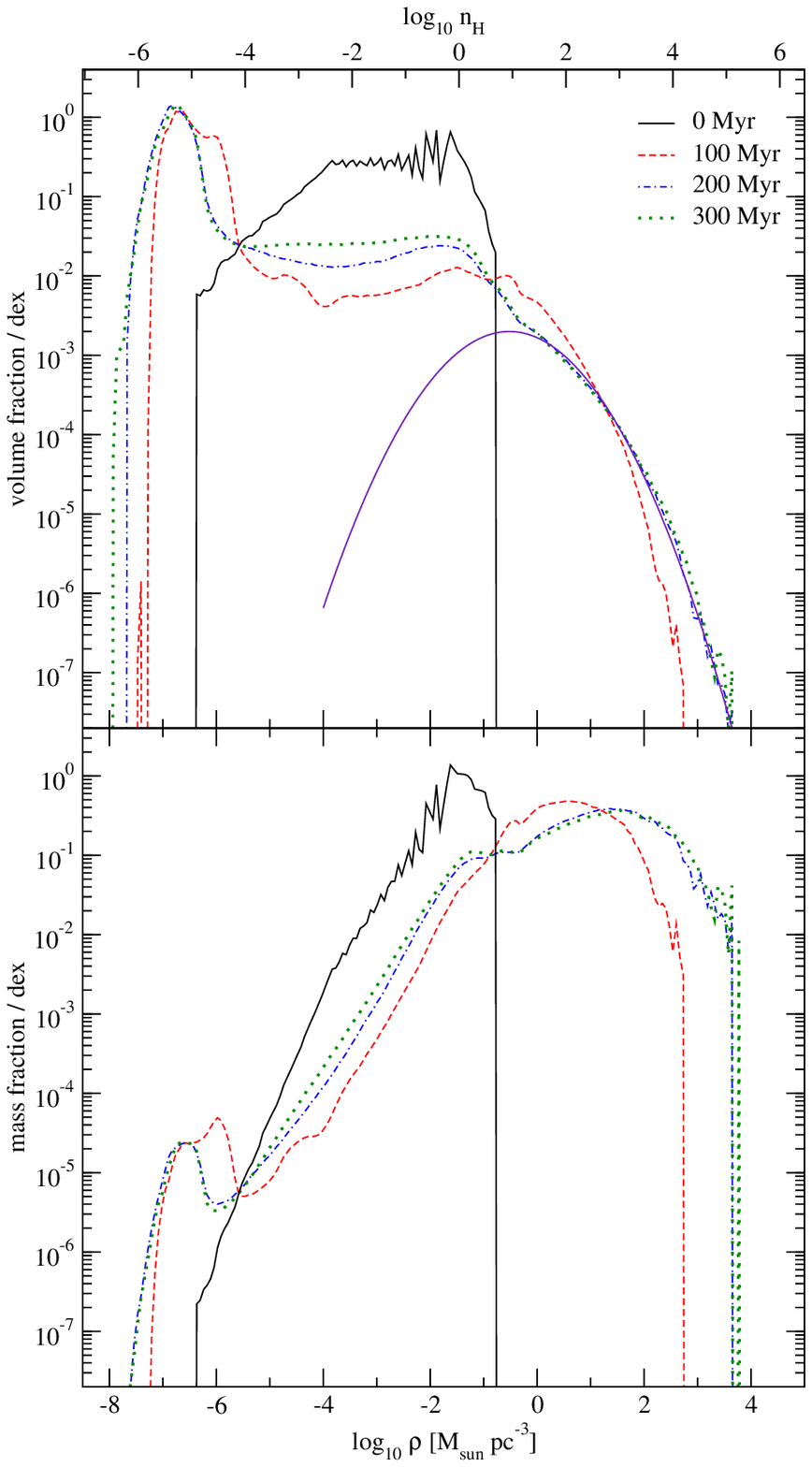}
\caption{Top panel: Evolution of volume-weighted probability
distribution function (PDF) for gas density evaluated from $2.5~{\rm
kpc}<r<8.5~{\rm kpc}$ and $-1~{\rm kpc}<z<1~{\rm kpc}$. The jagged
nature of the 0~Myr PDF is due to finite root grid resolution aliasing
of the initial conditions, and does not influence the later
evolution. Higher and lower density regions develop due to
gravitational fragmentation. Note there is relatively little evolution
between 200 and 300~Myr. The volume fractions of GMCs (i.e. cells with
$n_{\rm H}>100\:{\rm cm^{-3}} = 3.46 M_\odot\:{\rm pc^{-3}}$) at
$t=100,200,300$~Myr are $(5.82, 4.28, 4.12)\times 10^{-4}$,
respectively. The smooth solid line is a log-normal fit to the portion of the PDF at cloud densities (see text). Bottom panel: Mass-weighted PDF for gas density
evaluated over the same simulation volume. Again, these distributions
show relatively little evolution between 200 - 300\,Myr, and the GMC
mass fractions at $t=100,200,300$~Myr are 0.492, 0.688, 0.685,
respectively.
\label{fig:disk_pdf}}
\end{center} 
\end{figure}


The evolution of the disk is shown in
Figures~\ref{fig:disk_evol_images} \& \ref{fig:disk_evol_plots}.
Starting with a velocity dispersion of $\sigma_g = c_s = 9$\,km~s$^{-1}$,
(i.e. at a temperature of 7450\,K) the disk is gravitationally
stable. The gas cools relatively quickly, lowering the sound speed to
below 6~km~s$^{-1}$, at which point $Q<1$ in the main disk region. This
cooling occurs most rapidly in the denser inner regions.

The first structure to form is an overdense ring near the inner
boundary of the main disk at $r=2$~kpc. The formation of this
structure is influenced by the boundary condition of the inner edge of
the main disk: i.e. the relative lack of gas inside 2~kpc. The Toomre
ring instability gathers material radially from a scale about equal to
the Toomre length, $\lambda_T = 2\pi^2G\Sigma_g/\kappa^2$ (the numerical
coefficient applies for infinitely thin gas disks), i.e. the most
unstable scale. Similar ring structures form slightly later just
outside the inner ring and also at the outer edge of the disk. These
rings then fragment azimuthally into clouds. To avoid structures being
influenced by the boundaries of the main disk, we restrict our
cloud analysis to clouds that form between 2.5-8.5~kpc.

The fragmentation of the main disk begins at the inner part of this
region, not only because the cooling occurs faster in this denser
region, but also because once gravitationally unstable structures have
formed, being denser they collapse with shorter free-fall times.  By
$t\sim 140$~Myr the main region of the disk, i.e. out to $\sim
8.5$~kpc, has fully fragmented.

Once formed, clouds start to interact as differential rotation in the
disk brings them into contact with one another, and again this
evolution occurs more rapidly in the inner regions of the
disk. Gammie, Ostriker, \& Jog (1991) investigated this process
analytically and with numerical integrations of binary collisions and
cloud interactions, finding interaction times that are a fraction of
the orbital time. We investigate the cloud collision time in more
detail in \S\ref{sec:cloud_merger}.

Outside the main disk region, the low density gas remains stable. The
very center of the disk, $r < 2$\,kpc, does show some development of
overdense structures. This is due to the infall of gas to the galactic
center, most likely due to numerical viscosity resulting from the use
of a Cartesian grid to simulate small-scale circular gas
motions. These inner regions are excluded from our analysis.

Azimuthally-averaged radial profiles of $\Sigma_g$, $\sigma_g$, $T$,
and $Q$ are shown in Figure~\ref{fig:disk_evol_plots}.  Note that here
we show $\Sigma_g = \int_{-1{\rm kpc}}^{+1{\rm kpc}}\rho(z) d z$,
which is effectively equal to the full vertical mass surface density
through the simulation box, $\sigma_g$ is a mass-weighted average over
$-1~{\rm kpc}<z<1~{\rm kpc}$ utilizing only disk plane velocity
components, $T$ is a mass-weighted average over $-1~{\rm kpc}<z<1~{\rm
kpc}$ and $Q$ makes use of $\Sigma_g$ and $\sigma_g$ via equation
(\ref{Q}).


In the main disk, away from the boundaries, the mean value of
$\Sigma_g$ at a particular radius does not change very much during the
course of the simulation. Small scale fluctuations due to ring
instabilities are prominent at early times, but become smoothed out as
the rings fragment into clouds, which then gravitationally scatter off
of each other.


During the early stage of the simulation, the gas cools rapidly,
which causes $Q$ to drop below the threshold for instability.
The gas is also contracting vertically towards the disk midplane.
After about 50\,Myr, the bulk properties of the main disk remain
relatively constant for some time, since the temperature floor of
300~K has been reached for much of the disk material. At later times,
after the disk has fragmented into clouds and they start interacting
gravitationally, there is a significant increase in the velocity
dispersion of the gas and some heating. At late times the Toomre
stability parameter rises to values above unity, because of the strong
gravitational scattering of clouds.

\begin{figure*} 
\begin{center} 
\includegraphics[width=\textwidth]{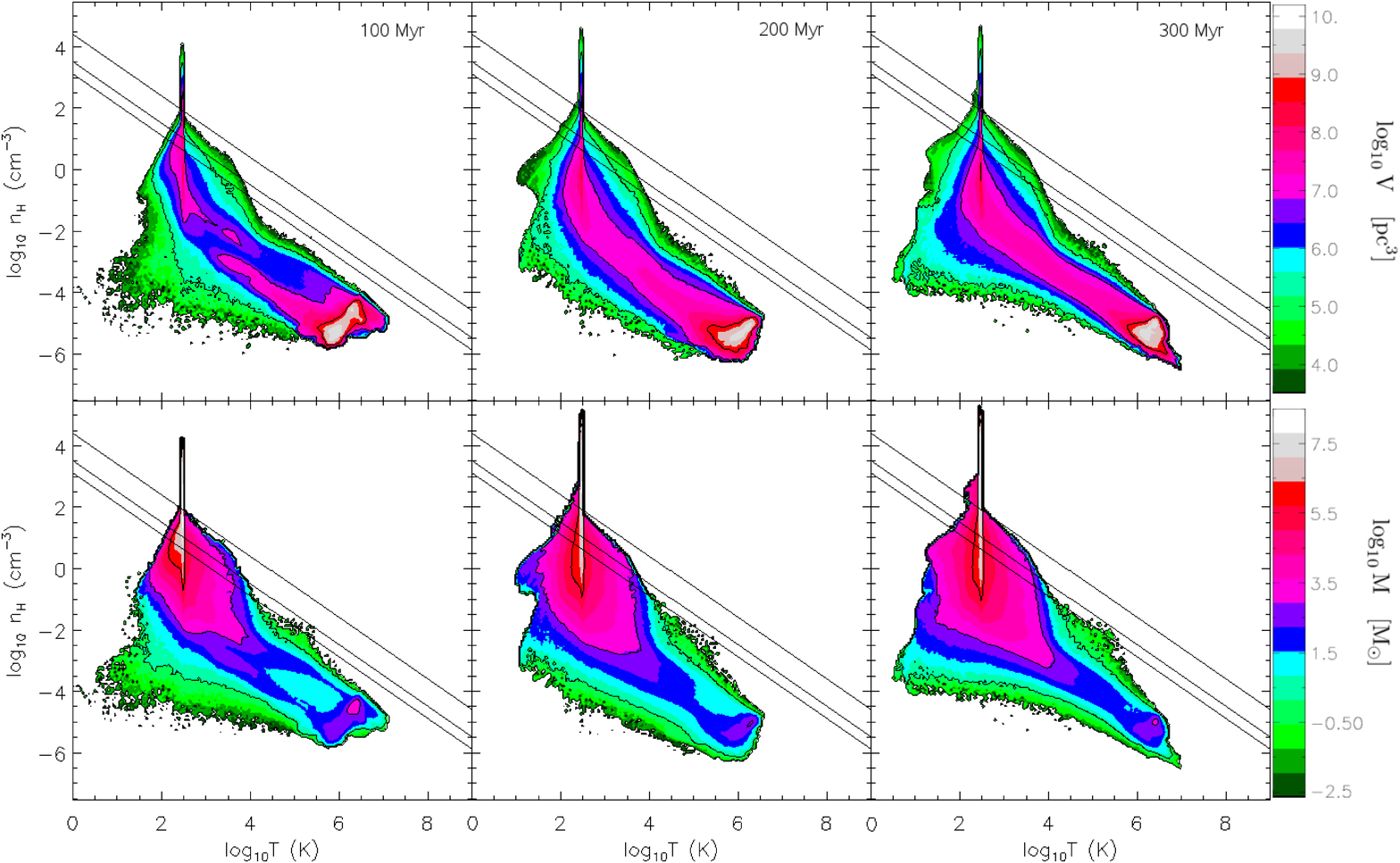}
\caption{Density versus temperature contour plots showing the
distribution and evolution (100, 200, 300~Myr from left to right) of
gas volume (top row) and mass (bottom row) in the galaxy disk for
$2.5~{\rm kpc}<r<8.5~{\rm kpc}$ and $-1~{\rm kpc}<z<1~{\rm kpc}$. The
temperature floor of 300~K is evident: most ``GMC'' gas has this
effective temperature, corresponding $c_s\simeq 1.8\:{\rm
km\:s^{-1}}$. Lower temperatures are possible via adiabatic
cooling. Most of the volume of the ISM exists at lower densities and
higher temperatures, with different phases in approximate pressure
equilibrium. In each panel, solid lines show estimates of the total
pressure in the Milky Way, $P_{\rm tot}/k= 2.8 \times 10^4$\,K
cm$^{-3}$ (top line), the total thermal pressure, $P_{\rm th}/k =
0.36 \times 10^4$ K cm$^{-3}$ (middle line), and the thermal pressure
excluding the hot gas component, $P_{\rm th, no hot}/k = 0.14 \times
10^4$ K cm$^{-3}$ (bottom line) (Boulares \& Cox 1990). Since our
simulations do not yet include feedback (e.g., FUV heating,
supernovae) or nonthermal pressure components (e.g., magnetic fields,
cosmic rays), it is not surprising that our diffuse ISM pressure is
1-2 orders of magnitude smaller than the observed Milky Way
value. 
\label{fig:disk_contours}}
\end{center} 
\end{figure*}

Figure~\ref{fig:disk_pdf} shows the probability distribution function
(PDF) for gas density. The top panel shows the volume-weighted PDF,
evaluated over a volume extending radially from 2.5 to 8.5~kpc and
$\pm1$~kpc above and below the disk midplane.  The mass-weighted PDF
is shown in the bottom panel. These figures show the relatively fast
evolution from the initial conditions caused by the early cooling and
fragmentation. Evolution after 100~Myr proceeds more slowly: there is
very little change from 200 to 300~Myr.

Figure~\ref{fig:disk_pdf} also shows a fit of a log-normal
distribution, $ p({\rm ln}x){\rm d\:ln}x = (2\pi \sigma^2_{\rm
  PDF})^{-1/2} {\rm exp}(-0.5 \sigma_{\rm PDF}^{-2}[{\rm ln}x -
  \overline{{\rm ln}x}]^2)$, where $x=\rho/\overline{\rho}$, to the
volume-weighted PDF at the densities relavant to clouds (see also Wada
\& Norman 2007; Tasker \& Bryan 2008). Since we are only fitting to a
portion of the PDF, here we are only interested in the width of the
distribution, $\sigma_{\rm PDF}$, not the normalization. We find
$\sigma_{\rm PDF}=2.0$. Following the empirical relation $\sigma_{\rm
  PDF}^2 = {\rm ln}[1+(3 {\cal M}^2/4)]$ derived from analysis of
simulations of isothermal, non-self-gravitating supersonic turbulence
(Padoan, Nordlund \& Jones 1997, Padoan \& Nordlund 2002, Krumholz \&
McKee 2005), where ${\cal M}$ is the 1D Mach number, we estimate
${\cal M} = 8.5$. For a sound speed of $1.80\:{\rm km\:s^{-1}}$, this
corresponds to a velocity dispersion of $15\:{\rm km\:s^{-1}}$, about
50\% larger than the typical internal velocity dispersions of clouds
(\S\ref{sec:cloud_simtime}) or the disk-mass-averaged velocity
dispersions (Fig.~\ref{fig:disk_evol_plots}). This moderate
discrepancy may be due to self-gravity skewing the high-side of the
PDF and/or the effects of shearing streaming motions in the disk,
which are removed from the disk-averaged velocity dispersions.



The density-temperature phase space of the ISM is shown in
Figure~\ref{fig:disk_contours}. In the top row, the contours are
related to the volume in the simulation at the given densities and
temperatures. After disk fragmentation, most of the volume is at low
densities, $n_{\rm H} \sim 10^{-5}\:{\rm cm^{-3}}$, and high
temperatures, $T\sim 10^6\:{\rm K}$. The temperature floor of the
cooling curve at 300~K is evident on the left hand side of these
diagrams: most of the GMC material is at this {\it effective}
temperature, i.e. has an effective sound speed of 1.8~km~s$^{-1}$, and
these clouds occupy very little volume. Note that cooler temperatures
are possible via adiabatic cooling. In the bottom row, the contours
are related to the mass in the simulation at the given densities and
temperatures. Most mass is in high density, $n_{\rm H} \sim
1-1000\:{\rm cm^{-3}}$, structures, including our defined ``GMCs''
with $n_{\rm H} \geq 100\:{\rm cm^{-3}}$.

The typical local Milky Way total diffuse ISM pressure is about
$2.8\times 10^4\:{\rm K~cm^{-3}}$ and its thermal components are about
an order of magnitude smaller (Boulares \& Cox 1990). These pressure
are shown by straight lines in Figure~\ref{fig:disk_contours}. Our
simulated diffuse ISM is at significantly lower pressures compared to
the observed Milky Way pressures. This is not surprising since this
simulation does not include feedback from star formation, including
FUV heating, stellar winds, ionization and supernovae. Nevertheless,
much of the volume of the simulated ISM is in approximate pressure
equilibrium. The pressure is set by energy input from hot gas produced
in shocks resulting from cloud-cloud collisions. GMCs in the
simulation are at much higher pressures than the diffuse ISM, due to
their self-gravity (see below). In fact the thermal pressure of the
cloud threshold density at the minimum cooling temperature is about
equal to the mean total pressure in the local Milky Way ISM observed
by Boulares \& Cox (1990).

\section{GMC Properties and Evolution}

\subsection{GMC Formation and Merger Rates}
\label{sec:cloud_merger}

\begin{figure} 
\begin{center} 
\includegraphics[width=\columnwidth]{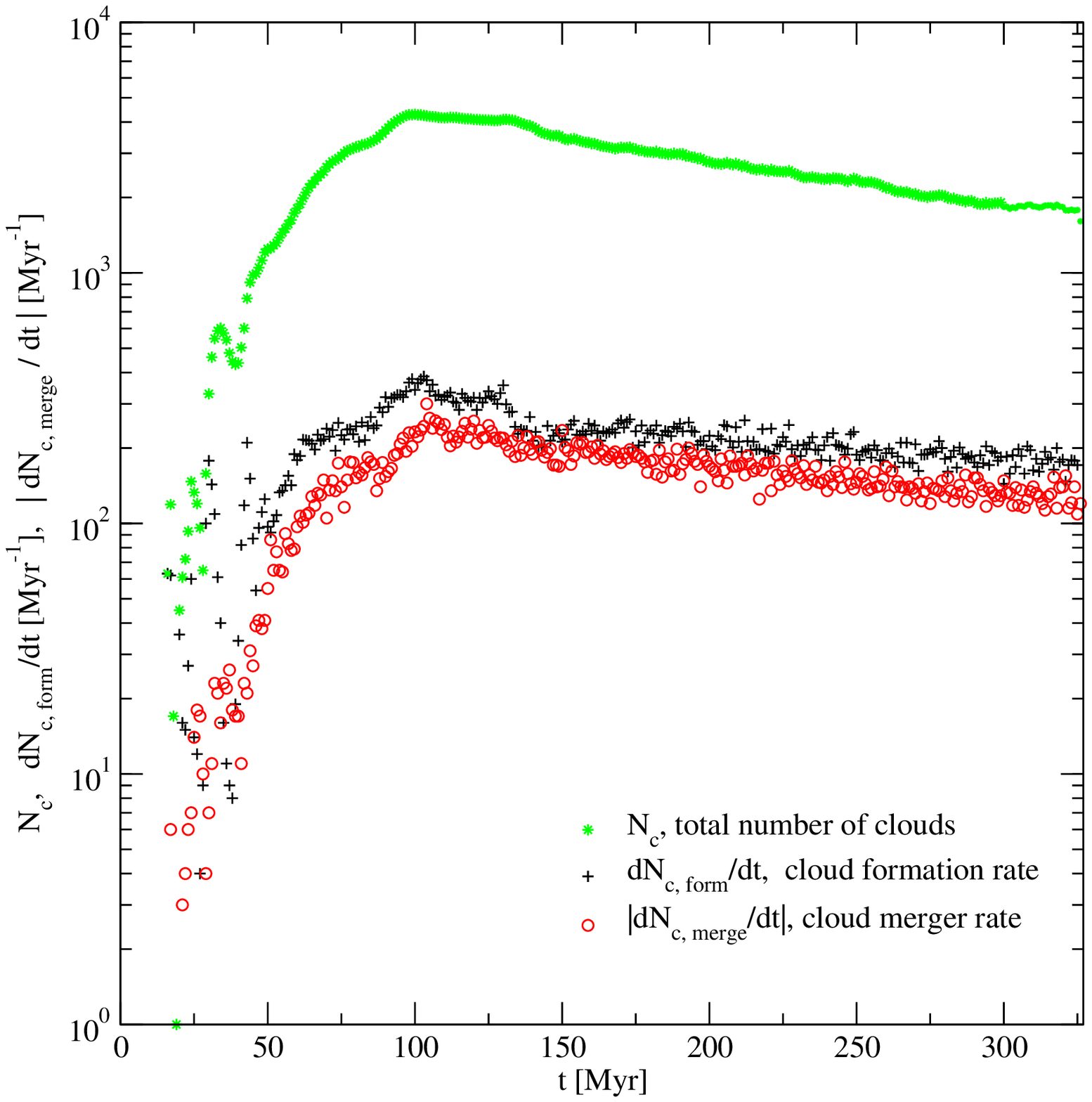}
\caption{Evolution of the number of GMCs, $N_c$, the GMC formation
rate, $dN_{\rm c,form}/dt$, and the GMC merger rate, which is
essentially equal in magnitude to the rate at which clouds are
destroyed via merging, $dN_{\rm c,merge}/dt$, since virtually all
mergers are binary mergers. The cloud formation rate shows a burst at
about 100~Myr associated with the initial fragmentation of the disk,
which then decreases by a factor of about two by 140~Myr, after which it
remains almost constant for the remainder of the simulation. Note that
clouds can also be destroyed by disruptive collisions and by leaving
the (inner) boundary of the analyzed volume. Thus in the
fully-fragmented phase ($t>140$~Myr) the overall cloud destruction
rate is greater than the formation rate by about 10 clouds per Myr.
}
\label{fig:formation_history}
\end{center} 
\end{figure}

\begin{figure} 
\begin{center} 
\includegraphics[width=\columnwidth]{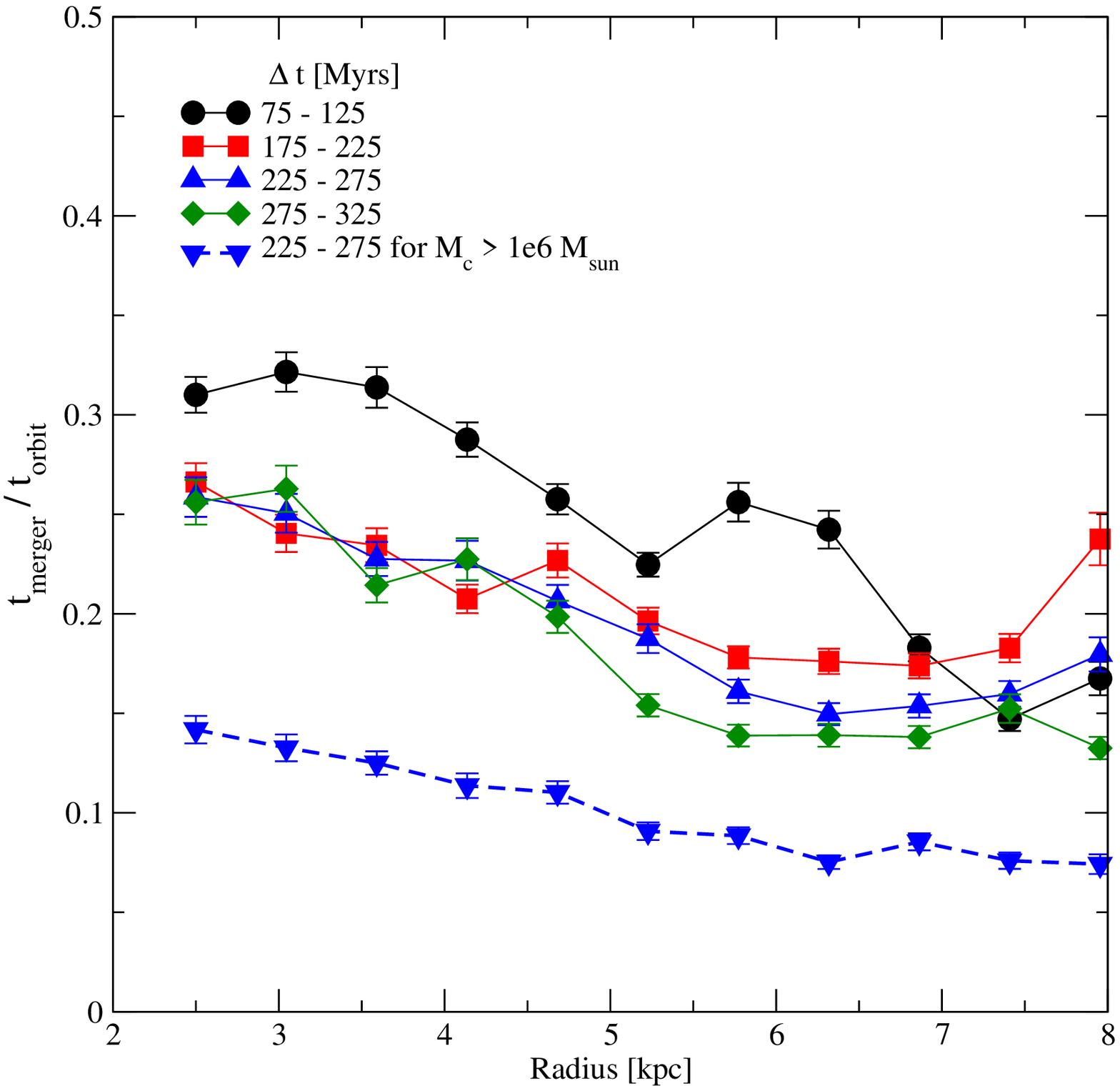}
\caption{Cloud merger timescales (averaged over 50~Myr intervals of
  simulation time) compared to orbital timescale as a function of
  galactocentric radius. The merger rate is driven by differential
  rotation in the disk, and the cross-section grows as clouds become
  more gravitationally bound. After the initial fragmentation stage at
  100~Myr, the cloud merger timescale settles to small values
  $\simeq0.2t_{\rm orbit}$, with only modest dependence on
  galactocentric radius. The dashed line shows the merger times of
  clouds with $M_c>10^6M_\odot$ (i.e. the average time for them to
  undergo a merger with a cloud of any mass, though typically with $M_c>10^5M_\odot$ [see Fig.~\ref{fig:simtime_resolution}a]) evaluated over the
  interval $t=225-275$~Myr.}
\label{fig:merger_rate}
\end{center} 
\end{figure}


As the disk becomes gravitationally unstable, local patches increase
in density, many of which reach our threshold value of $n_{\rm H,c} =
100$\,cm$^{-3}$, at which point we recognize them as a GMC.  The
formation rate of these clouds over the course of the simulation is
shown in Figure~\ref{fig:formation_history}. There is a burst of cloud
formation associated with the initial fragmentation of the disk, with
the rate peaking at about $400\:{\rm Myr}^{-1}$ at a simulation time
of $t=100$~Myr. By $t=140$~Myr, cloud formation settles down to a
slower and nearly constant rate of $\sim 200\:{\rm Myr}^{-1}$. This is
partly due to there being a smaller non-cloud diffuse gas reservoir
and partly due to heating of the disk by gravitational interactions of
the cloud population. We are most interested in the cloud and ISM
properties after $t = 140$~Myr, since before this time their
properties are influenced by the artificially smooth initial
conditions and their fragmentation via the Toomre ring instability.

Figure~\ref{fig:formation_history} also shows the evolution of the
total number of clouds and the merger rate of clouds. This information
is used to calculate the mean merger time of clouds, $t_{\rm merger}$,
relative to their orbital time, as a function of galactocentric radius
(Figure~\ref{fig:merger_rate}). This calculation assumes mergers are all
binary mergers (the non-binary merger fraction is indeed negligibly
small). From Figure~\ref{fig:merger_rate} we see that the average merger
time settles to be a small fraction, $\sim 0.2$ of the orbital time,
with only a modest dependence on galactocentric radius and simulation
time for $t\geq 140$~Myr. This confirms the results of Tan (2000), who
estimated $t_{\rm merger}/t_{\rm orbit}=0.2$ based on the simplified
calculation of cloud orbits and binary interactions by Gammie et al. (1991). Note that the
orbital time for $v_c=v_{c,0}=200\:{\rm km\:s^{-1}}$ is
\begin{equation}
\label{torbit}
t_{\rm orbit} = 123 (r/4\:{\rm kpc}) (v_c/200\:{\rm km\:s^{-1}})^{-1} ~{\rm Myr},
\end{equation}
so that the typical merger time is only $\sim 25$~Myr, which is
relatively short compared to traditional estimates of cloud collision
time scales of hundreds of Myrs that do not allow for the essentially
2D geometry of the GMC population, gravitational focusing and that
cloud interaction velocities are set by galactic shear and are larger
(by about a factor of two) than the local cloud velocity dispersion
(see discussion in McKee \& Ostriker 2007). Our derived collision
timescales are also shorter than many estimates of GMC lifetimes
(e.g. McKee \& Williams 1997; Matzner 2002), which suggests that
collisions are likely to be important even when feedback mechanisms
are taken into account: i.e. an individual GMC is just as likely to
have its properties dramatically altered by a merger than by a
destructive mechanism such as supernova or ionization feedback.

The nearly constant ratio of $t_{\rm merger}/t_{\rm orbit}$ means that
the cloud collision rate is tied to the global galactic dynamical
timescale and, if cloud collisions are the trigger for the majority of
star formation, then this provides a natural mechanism to explain the
global SFR-gas content correlations observed by Kennicutt (1998) in
which the overall star formation efficiency per galactic dynamical
time is small, while having star formation occur mostly on small
scales at relatively high efficiency in a highly clustered mode (Tan
2000). The relation of star formation to cloud collisions in global
galaxy simulations will be investigated in Paper II.

The following caveats should be considered in the above estimates of
the merger timescale. The simulated merger rate depends on the mass
function and gravitational boundedness of the GMCs. More massive and
more gravitationally bound GMCs will experience higher merger rates,
although for typical cloud mass functions most of their collisions
will be with lower-mass clouds. In Figure~\ref{fig:merger_rate} we
also show the merger timescale for clouds with $M_c\geq10^6M_\odot$,
which is about a factor of two shorter than that of the average cloud.
We will see below that the simulated GMC population mass function and
gravitational boundedness are similar to observed values. The total
number of clouds is not fully resolved in these simulations (the most
common cloud mass is several$\times 10^5\:M_\odot$ in the high
resolution run at late times). Higher resolution simulations are
likely to resolve larger number of clouds, which would increase the
merger rate, although not necessarily the total gas mass that has been
compressed by collisional shocks. The inclusion of stellar feedback
would likely lead to a reduced total mass fraction in clouds and a
reduced gravitational boundedness of those clouds that do form, and
these effects would reduce the merger rate.


\subsection{GMC Properties with Simulation Time}
\label{sec:cloud_simtime}

\begin{figure*} 
\begin{center} 
\includegraphics[width=15cm]{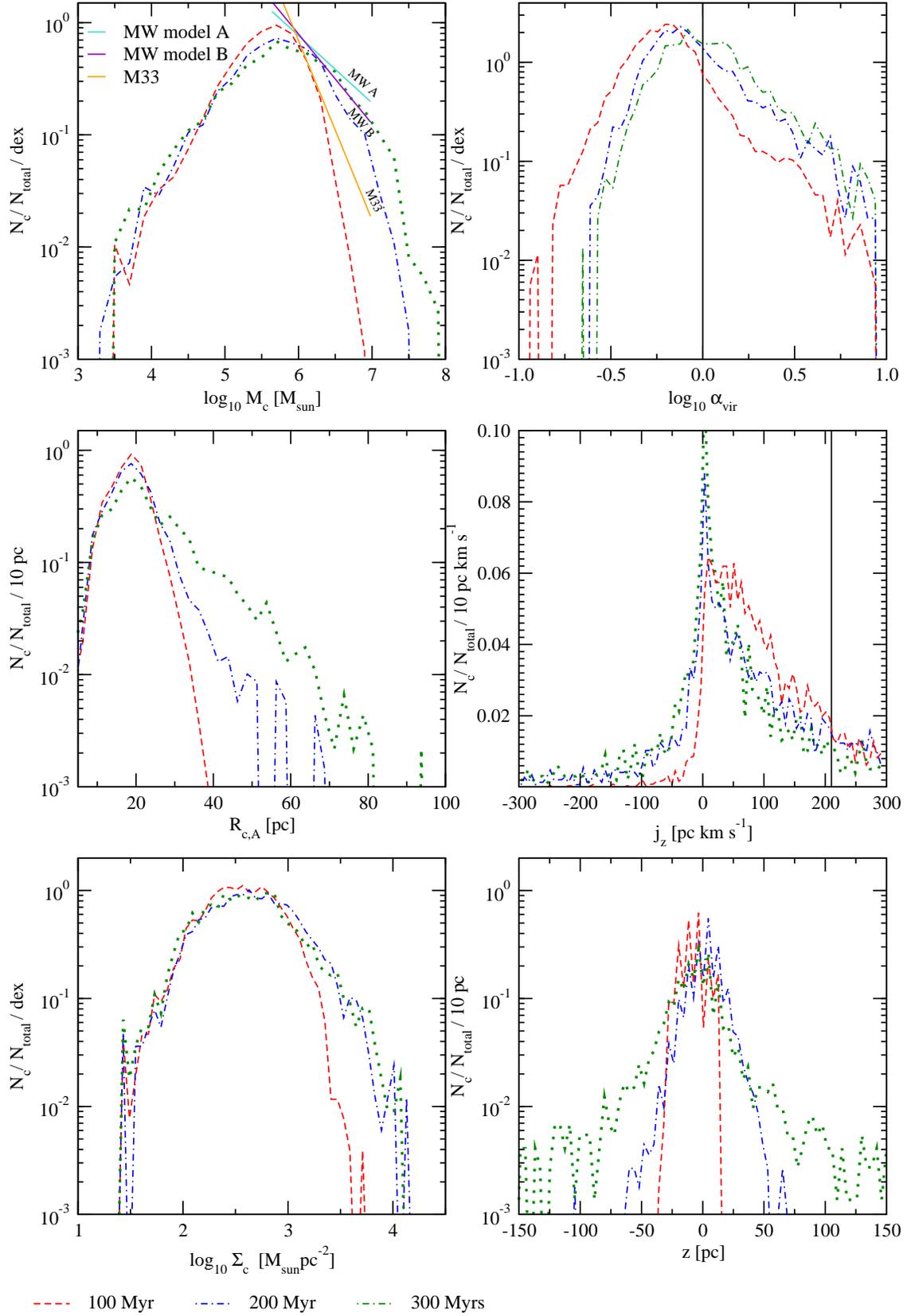}
\caption{Normalized distributions of GMC properties at 100 (dashed
lines), 200 (dot-dashed lines) and 300 (dotted lines)\,Myr after the
start of the simulation, for which the total number of clouds are
4300, 2770, 1840, respectively. Top left: (a) Cloud mass, $M_c$. The
total number of clouds with $M_c\geq 10^6\:M_\odot$ are 625, 856, 694
for $t=100,200,300$Myr, respectively. Also shown are example
observational results for the power law slopes of GMC mass functions
in the Milky Way (Williams \& McKee 1997) and M33 (Rosolowsky et
al. 2003). Middle left: (b) Cloud radius, $R_{c,A} \equiv
(A_c/\pi)^{1/2}$, where $A_c$ is the projected area of the clouds in
the Y-Z plane. The minimum size cloud with 1 cell has
$R_{c,A}=4.4$~pc. Bottom left: (c) Mass surface density,
$\Sigma_c=M_c/A_c$. Top right: (d) Virial parameter, $\alpha_{\rm
vir}$. The vertical line indicates $\alpha_{\rm vir}=1$. Middle
right: (e) Vertical ($z$) component of specific angular momentum,
$j_z$. The vertical line indicates the value of $j_z$ of a spherical
($\simeq 110$~pc radius) region of the initial conditions at
galactocentric radius $r=4$~kpc containing $10^6\:M_\odot$. Bottom
right: (f) Cloud center of mass vertical positions. The RMS
heights of the cloud population at 100, 200, 300~Myr are 13, 25 and
51~pc, respectively.}
\label{fig:simtime_properties}
\end{center} 
\end{figure*}

\begin{figure*} 
\begin{center} 
\includegraphics[width=15cm]{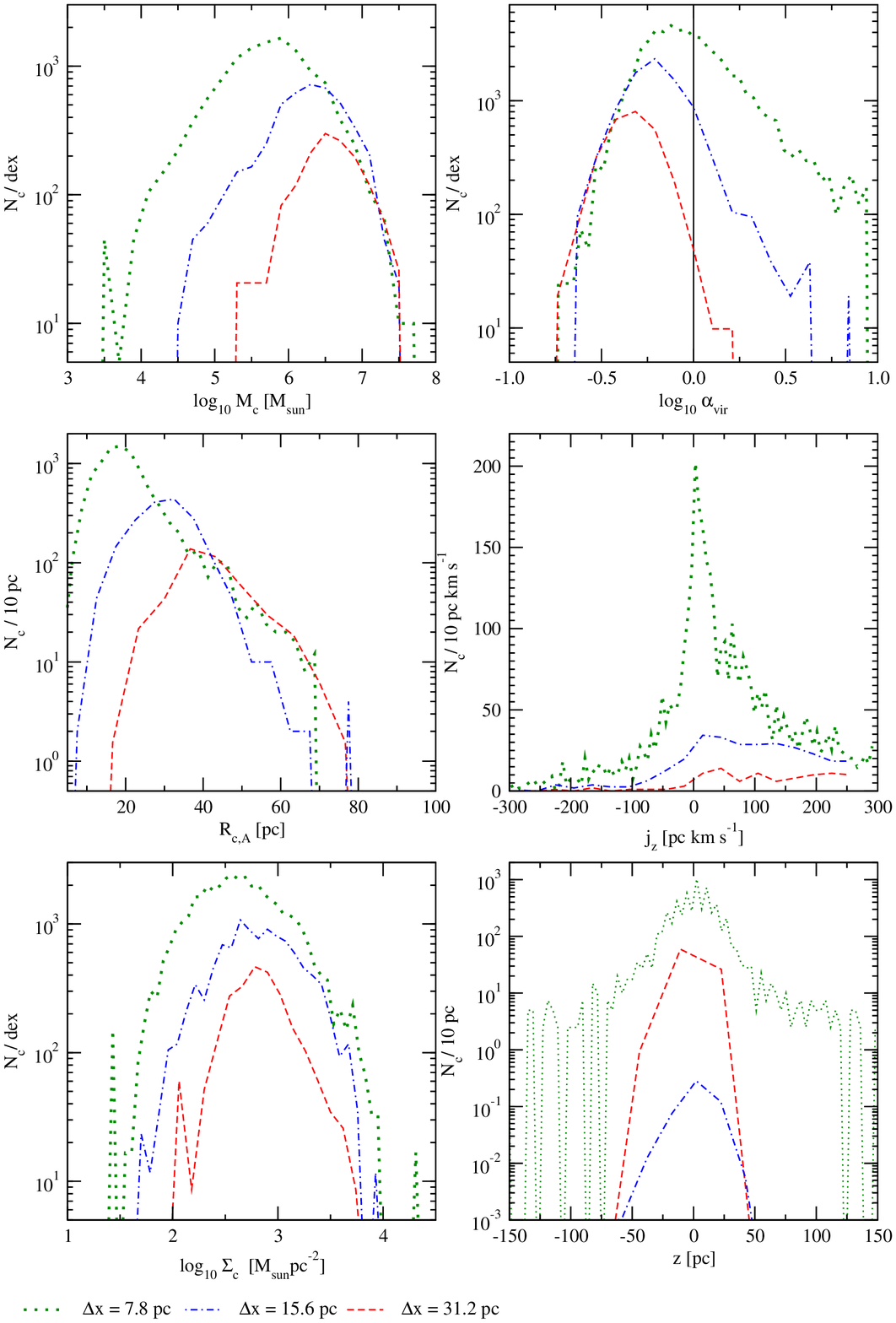}
\caption{Dependence of un-normalized distributions of GMC properties at
  $t=$250 Myr with minimum simulation resolution: 31.2~pc (dashed
  lines), 15.6~pc (dot-dashed lines) and 7.8~pc (dotted lines), for
  which the total number of clouds are 280, 884, 2366, respectively.
Top left: (a) Cloud mass, $M_c$. Middle left: (b) Cloud radius, $R_{c,A} \equiv
(A_c/\pi)^{1/2}$, where $A_c$ is the projected area of the clouds in
the Y-Z plane. Bottom left: (c) Mass surface density,
$\Sigma_c=M_c/A_c$. Top right: (d) Virial parameter, $\alpha_{\rm
vir}$. The vertical line indicates $\alpha_{\rm vir}=1$. Middle
right: (e) Vertical ($z$) component of specific angular momentum,
$j_z$. 
Bottom
right: (f) Cloud center of mass vertical positions.}
\label{fig:simtime_resolution}
\end{center} 
\end{figure*}

The cloud mass function is shown in
Figure~\ref{fig:simtime_properties}a. Since there is no star formation
feedback in these models, it is difficult to destroy gravitationally
bound gas clouds, except via their merger into more massive clouds or
via disruptive collisions. Unbound and weakly bound clouds can also be
destroyed by shearing tidal forces due to grazing and close cloud
interactions.  As a result, we expect the mean cloud mass to grow over
the course of the simulation, and this behavior is seen in
Figure~\ref{fig:simtime_properties}a. Note, however, that the peak of
the cloud mass function is at $6\times 10^5\:M_\odot$ (for uniform
binning in ${\rm log} M_c$) and this does not vary significantly from
$t=$100 to 300~Myr. Note the initial Toomre mass in a $Q=1$ disk is
$M_T \equiv \Sigma_g \lambda_T^2 =8.9\times 10^6 (\sigma_g/6{\rm
  km\:s^{-1}})^3 (r/4{\rm kpc}) (v_c/200{\rm km\:s^{-1}})^{-1}
M_\odot$, about an order of magnitude larger than the peak of the
simulated GMC mass function. The Jeans mass in the disk is $M_J = \pi
\Sigma_g \lambda_J^2 \sigma_g^4/(G^2\Sigma_g)\rightarrow 2.2\times
10^6 (\sigma_g/6){\rm km\:s^{-1}})^3 (r/4{\rm kpc}) (v_c/200{\rm
  km\:s^{-1}})^{-1} (Q/1)\:M_\odot$, where $\lambda_J \equiv
\sigma_g^2/(G\Sigma_g)$ is the shortest wavelength permitting
gravitational instability in a thin, nonrotating disk (Kim \& Ostriker
2001). The gas tends to cool to the temperature floor of 300~K, for
which $\sigma_g \simeq 2\:{\rm km\:s^{-1}}$ and thus $Q\simeq 1/3$,
resulting in $M_J \simeq 3\times 10^4\:M_\odot$.  The effect of
simulation resolution on the cloud mass function at $t=250$~Myr is
shown in Figure~\ref{fig:simtime_resolution}a. The number of GMCs
appears reasonably well-resolved down to masses of about $2\times
10^6\:M_\odot$.


The simulated cloud mass function can be compared to the observed mass
spectrum of GMCs in the Milky Way and M33, which can be fitted by a
power law $d{\cal N}_c /(d {\rm ln} M_c) \propto
M_c^{-\alpha_c}$. In the Milky Way, $\alpha_c$ is observed to be
$\simeq 0.6-0.8$ (Williams \& McKee 1997), whereas a steeper index of
$\simeq 1.6$ appears to hold in M33 (Rosolowsky et al. 2003).
After collision and agglomeration processes have had enough
time to operate, i.e. by $t=200$~Myr, the shape of the simulated cloud
mass function (in the mass range $\sim (0.5-10)\times 10^6\:M_\odot$
relevant to observations) approaches that seen in the Milky Way GMC
population, indicating that most of the total cloud mass is contained
in the more massive clouds.  It remains to be determined whether these
processes would still be as important for shaping the cloud mass
function if stellar feedback processes were also operating, which
would tend to reduce cloud lifetimes. 

Williams \& McKee (1997) also derived the normalization of the
Galactic GMC mass function and noted there appears to be a truncation
above a molecular mass of $M_u=6 \times 10^6$\,M$_\odot$ (see
eq.~\ref{cloud_mf}). Allowing for an equal amount of atomic gas
associated with GMCs (Blitz 1990), this truncation would occur at
$1.2\times 10^7\:M_\odot$. We see that at late simulation times, the
cloud mass function has grown beyond this value, though there are very
few clouds with $M_c>10^7\:M_\odot$ (see also
Figure~\ref{fig:simtime_resolution}a). The presence of these very
massive clouds is likely due to the lack of stellar feedback cloud
destruction mechanisms in these current simulations, especially the
ability of stars to destroy their natal clouds. Williams \& McKee
(1997) estimate there are $\sim 100-200$ inner Milky Way GMCs with
$M_c>10^6\:M_\odot$, and about 1000 with $M_c>10^5\:M_\odot$. If one
allows for an equal mass of atomic gas associated with the observed
molecular gas of these GMCs, then in comparison we find about 400
clouds with $M_c>2\times 10^6\:M_\odot$, a factor of 2 to 4 higher
than the Williams \& McKee estimate. We do not adequately resolve the
number of $2\times 10^5\:M_\odot$ clouds
(Figure~\ref{fig:simtime_resolution}a) to make a useful comparison at
that mass scale. These results may indicate that the number of massive
GMCs in our simulation is a factor of a few larger than in the Milky
Way. Indeed the mass fraction of gas in GMCs in the simulation is
about 0.69 (Figure~\ref{fig:disk_pdf}), somewhat higher than the mass
fraction implied by the analysis of Wolfire et al. (2003). Lack of
feedback processes in the simulation is an obvious potential cause of
this discrepancy. Another potential contributing factor is our
somewhat arbitrary choice of $n_{\rm H,c}=100\:{\rm cm^{-3}}$ for the
cloud threshold density. Note that M33, being a much smaller galaxy
than the Milky Way, has fewer massive GMCs.

Figure~\ref{fig:simtime_properties}b shows the average radius of the
clouds defined as $R_{c,A} \equiv (A_c/\pi)^{1/2}$, where $A_c$ is the
projected area of a cloud in the y-z plane (i.e. as it would be
observed by an observer embedded in the plane of the galaxy).  The
most common radius for a cloud is around 20-30\,pc (for uniform linear
binning in $R_{c,A}$). By 300~Myr, agglomeration processes have
created a population of larger clouds. It is clear from
Figure~\ref{fig:simtime_properties}a \& b that the numbers of these
larger, more massive clouds are steadily increasing in time, and in
this respect a steady state has not been reached. Such a steady state
likely requires feedback processes. Nevertheless, the typical sizes of
the cloud population are similar to observed sizes of GMCs in the
Milky Way and other galaxies, such as M33 (Rosolowsky et al. 2003).
The effect of simulation resolution on the distribution of cloud sizes
is shown in Figure~\ref{fig:simtime_resolution}b.

Figure~\ref{fig:simtime_properties}c shows the distribution of mass
surface density of the clouds, which is defined as $\Sigma_c \equiv
M_c/A_c$. The peak of the distribution (for uniform binning in ${\rm
  log} M_c$) is at about $300\:M_\odot\:{\rm pc^{-2}}$, and does not
change significantly during the simulation. Although the number of
more massive clouds is increasing, these are also larger, and so have
similar values of $\Sigma_c$. Most clouds are within a factor of 3 of
this peak value. This surface density is similar to the mean value of
$\sim 200\:M_\odot\:{\rm pc^{-2}}$ derived by Solomon et al. (1987)
from a $^{12}$CO survey of the Galaxy. Heyer et al. (2008) have
derived smaller values $\sim 100\:M_\odot\:{\rm pc^{-2}}$ based on
better sampled $^{13}$CO surveys. To compare these observed values,
which are based solely on molecular line (CO) emission, with our
simulated clouds, one should also include their associated atomic gas,
which may contribute at about the factor of two level. We conclude
that, even without disruptive effects of star formation feedback, our
simulated GMCs attain values of $\Sigma_c$ that are similar to
observed values. This may indicate that the dominate source of
turbulent pressure support in GMCs is injected via cloud collisions
and interactions (i.e. turbulent converging flows) rather than via
star formation feedback. Note that Joung \& Mac Low (2006) found that
supernova driven turbulence was insufficient to explain the observed
ISM velocity dispersions.  The effect of simulation resolution on the
distribution of $\Sigma_c$ is shown in
Figure~\ref{fig:simtime_resolution}c: the peak of the distribution is
relatively insensitive to resolution. 

Figure~\ref{fig:simtime_properties}d shows the distribution of GMC
virial parameters (see eq.~\ref{alpha_vir}), $\alpha_{\rm vir} \equiv
5 \sigma_{c}^2 R_{c,A} / (GM_c)$, where $\sigma_{c}$ is the
mass-averaged 1D velocity dispersion of the cloud,
i.e. $\sigma_{c}\equiv (c_s^2+\sigma_{\rm nt,c}^2)^{1/2}$, where
$\sigma_{\rm nt,c}$ is the 1D RMS velocity dispersion about the
cloud's center of mass velocity (Bertoldi \& McKee 1992). A virial
parameter of unity implies a spherical, uniform cloud with negligible
surface pressure and magnetic fields is virialized, so that its total
kinetic energy is half the magnitude of the gravitational
energy. Surface pressure terms and a centrally concentrated density
distribution tend to raise the value of $\alpha_{\rm vir}$
corresponding to virial equilibrium (see eq.~\ref{alpha_vir} and the
discussion above).
We see that the distribution of cloud virial parameters peaks at
$\alpha_{\rm vir}<1$, and decreases slightly as the simulation
progresses: the peak of the virial parameter distribution at 300~Myr
is about 0.6. This indicates that the clouds in the simulation are
typically gravitationally bound, perhaps somewhat more strongly bound
than observed GMCs, though there is still a significant population of
unbound clouds.  The effect of increasing simulation resolution on the
distribution of $\alpha_{\rm vir}$ is shown in
Figure~\ref{fig:simtime_resolution}d: the peak of the distribution
moves to larger values and the fraction of unbound clouds grows.


Figure~\ref{fig:simtime_properties}e shows the distribution of the
vertical ($z$) component of specific angular momentum, $j_z$, of the
clouds. The fraction of retrograde (i.e. $j_z<0$) clouds grows at late
times as clouds have had more opportunities to suffer mergers and
close interactions (see also Dobbs 2008). The vertical line indicates
the value of $j_z$ of a spherical ($\simeq 110$~pc radius) region of
the initial conditions at galactocentric radius $r=4$~kpc containing
$10^6\:M_\odot$. (Kim et al. 2003 derive an analytic expression for
$j_z\propto \Omega R_{\rm c,A}^2$ for 2D patches of galactic disks.)
We see that GMCs typically have much smaller amounts of angular
momentum compared to similar masses of diffuse gas that would be
spread out over larger scales and be more susceptible to galactic
differential rotation.  The effect of simulation resolution on the
distribution of $j_z$ is shown in
Figure~\ref{fig:simtime_resolution}e: at higher resolution it becomes
more sharply peaked towards small value of $j_z$.

Figure~\ref{fig:simtime_properties}f shows the distribution of cloud
center of mass vertical positions relative to the disk midplane,
i.e. where the vertical acceleration due to the galactic potential is
zero. The RMS heights of the cloud population at 100, 200, 300~Myr are
13, 25 and 51~pc, respectively, growing at late times because of
cloud-cloud gravitational scattering. Note Stark \& Lee (2005) derived
a vertical scale height of Milky Way GMCs of $\lesssim 35$~pc.
The effect of simulation resolution on the vertical distribution of clouds 
is shown in Figure~\ref{fig:simtime_resolution}f.



\begin{figure*} 
\begin{center} 
\includegraphics[angle=-90,width=\textwidth]{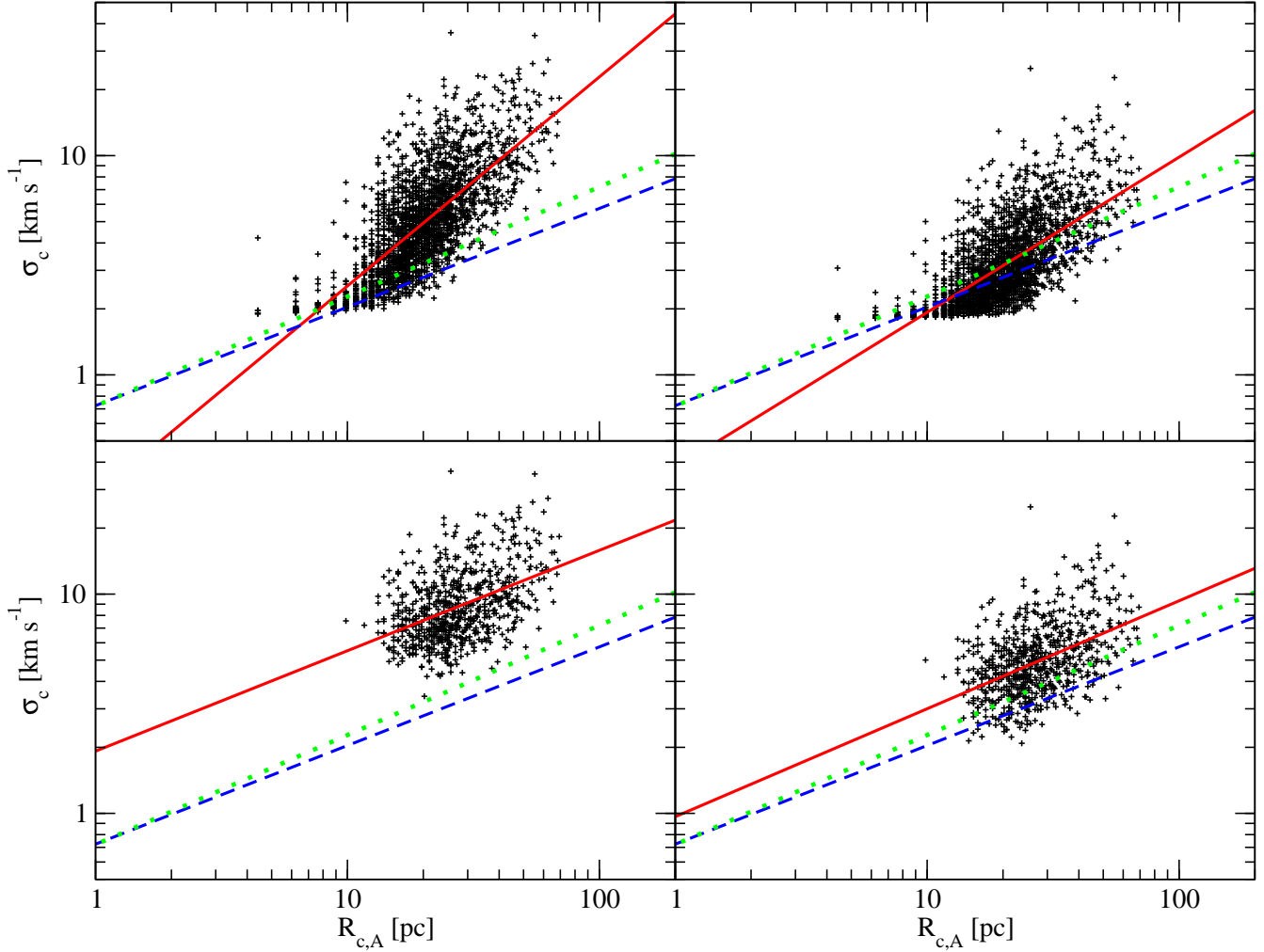}
\caption{Velocity dispersion versus size relation for clouds at
  $t=250$~Myr. The left panels show $\sigma_c$, the mass-averaged 1D
  internal velocity dispersion of clouds evaluated by averaging over
  the full 3D velocities. The right panels show $\sigma_c$ as it would
  be evaluated for clouds observed only with line of sight velocities
  in a galaxy viewed at $52^\circ$ inclination angle, similar to our
  view of M33. The top panels include all clouds, while the bottom
  panels include only clouds with $M_c\geq 10^6\:M_\odot$, i.e. those
  for which we have a better ability to resolve internal dynamics. The
  solid line shows the power law fits, $\sigma_c = \sigma_{\rm
    pc}(R_{c,A}/{\rm pc})^{\alpha_\sigma}$, to the simulated cloud
  population with power law slopes $\alpha_\sigma = 0.95, 0.71, 0.51,
  0.46$ (clockwise from top left) and $\sigma_{\rm
    pc}=0.283,0.378,0.97,1.92\:{\rm km\:s^{-1}}$, respectively. The
  dashed line shows the result of Rosolowsky et al.'s (2003) study of
  massive ($10^5 \:M_\odot \lesssim M_c\lesssim 10^6\:M_\odot$) GMCs in M33, with
  $\alpha_\sigma=0.45\pm0.02$ and $\sigma_{\rm pc}=0.72\:{\rm
    km\:s^{-1}}$. The dotted line shows the result of Solomon et al.'s
  (1987) study of Galactic GMCs, with $\alpha_\sigma=0.5\pm0.05$ and
  $\sigma_{\rm pc}=0.72\pm0.07\:{\rm km\:s^{-1}}$.
\label{fig:sigma_r}}
\end{center} 
\end{figure*}


\begin{figure*} 
\begin{center} 
\includegraphics[angle=270,width=\textwidth]{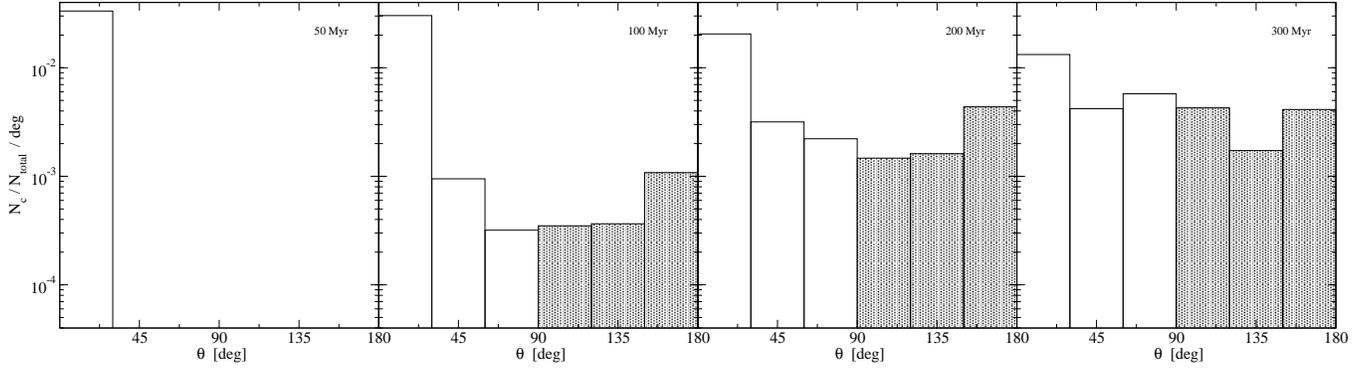}
\caption{Distribution of the angle, $\theta$, between cloud angular momentum
vectors and the galactic rotation axis at different times during the
course of the simulation. The shaded bars indicated retrograde
rotation, and this population grows with time as more and more clouds
experience collisions and close interactions.
\label{fig:cloud_theta}}
\end{center} 
\end{figure*}

\begin{figure*} 
\begin{center} 
\includegraphics[angle=270,width=\textwidth]{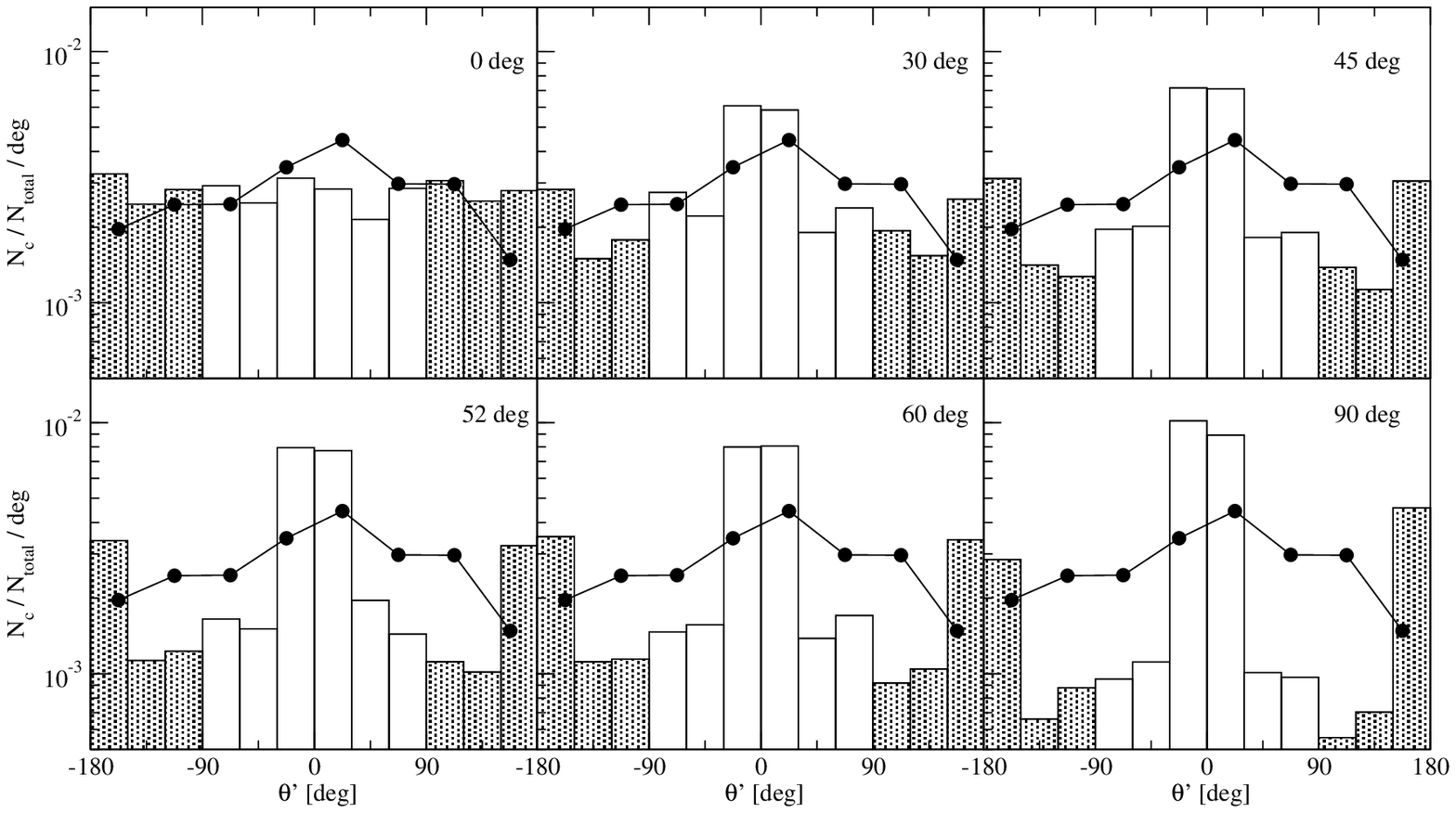}
\caption{Distribution of the position angle, $\theta^\prime$, of cloud
angular momentum vectors with respect to the galactic rotation axis as
determined from radial velocities for $t=250$~Myr, for which the
retrograde fraction of $\theta$ is 0.28, and for inclination angles of
0$^\circ$ (relevant to edge on disks including the Milky Way),
30$^\circ$, 45$^\circ$, 52$^\circ$ (relevant to M33), 60$^\circ$ and
90$^\circ$. Black line shows the observational results from
(Rosolowsky et al. 2003) for 45 clouds in M33 of which 18 are
retrograde (as defined by $\theta^\prime$).
\label{fig:cloud_thetaobs}}
\end{center} 
\end{figure*}

Figure~\ref{fig:sigma_r} shows the velocity dispersion versus size
relation for clouds at $t=250$~Myr. The effects of using the full
velocity information to derive a mass-averaged 1D internal velocity
dispersion versus just using radial velocities at a particular viewing
angle are examined. We also consider the effect of a mass cut,
i.e. restricting the analysis only to the most massive clouds. Such
effects are important in influencing the normalization and slope of
the line-width size relation. Our simulated cloud population has a
similar, though somewhat higher, velocity dispersion versus size
relation to observed GMCs, such as those observed in M33 by Rosolowsky
et al. (2003), in the Milky Way by Soloman et al. (1987), or the
average of a number of extragalactic systems, $\sigma_c = \sigma_{\rm
  pc}(R/{\rm pc})^{\alpha_\sigma}$ with
$\alpha_\sigma=0.60\pm0.10$ and $\sigma_{\rm pc}=0.44\pm0.15\:{\rm
  km\:s^{-1}}$, derived by Bolatto et al. (2008). The larger velocity
dispersion of the simulated clouds at a given size scale may be due to
the lack of star formation feedback causing them to be more tightly
gravitationally bound than real clouds. A discrepancy may also result
due to differences in how the effective radius of clouds is evaluated
for the simulation clouds and the real clouds. The dispersion of the
simulated clouds about the best fit relation is about $2.2\:{\rm
  km\:s^{-1}}$ (for the sample in the lower-right panel of
Figure~\ref{fig:sigma_r}), corresponding to about 40\% of
$\sigma_c$. Soloman et al. (1987) found a dispersion of about 30\% in
their sample of Galactic GMCs (see also Heyer \& Brunt 2004).

Figure \ref{fig:cloud_theta} shows the distribution of the angle,
$\theta$, between the cloud angular momentum vector and the galactic
rotation axis at different times in the simulation. At early times, in
the initial fragmentation phase clouds are all born with prograde
rotation ($0^\circ<\theta<90^\circ$), i.e. in the same sense as the
galactic rotation. At later times, cloud interactions lead to a build
up of a retrograde ($90^\circ<\theta<180^\circ$) cloud population,
which is about 30\% of the total.

To compare to observed systems, one needs to account for the viewing
angle to the galaxy and the fact that only line of sight velocities
are used to determine rotation. Figure~\ref{fig:cloud_thetaobs} shows
these effects by plotting the distribution of position angles,
$\theta^\prime$, of cloud rotation axes derived from radial velocities
along sight lines at different inclination angles to the simulated
galaxy at $t=250$~Myr, when the retrograde fraction of $\theta$ is
0.28. These results indicate that the derived retrograde fraction
depends on inclination angle: e.g. for the $0^\circ$ (edge on) case
the retrograde fraction infered from $\theta^\prime$ is about twice
that as measured by the actual angular distribution of $\theta$.  The
figure also shows the observational results of Rosolowsky et
al. (2003), who find about 40\% (18 out of 45) of the GMCs in M33 have
apparently retrograde rotation. Their results are broadly consistent
with our simulation results (note their slightly larger bin size), but
we emphasize that we can only make a qualititive comparison at this
stage, since our model galaxy was not set up to mimic the details of
M33 (e.g. its rotation curve), nor does it yet include the effects of
stellar feedback. Nevertheless cloud formation in a self-gravitating,
shearing disk and cloud evolution via cloud-cloud interactions appear
to be promising mechanisms to help explain the observed angular
distributions of cloud angular momentum vector orientations.


\subsection{GMC Properties with Cloud Age}
\label{sec:lifetimes}


\begin{figure*} 
\begin{center} 
\includegraphics[width=15cm]{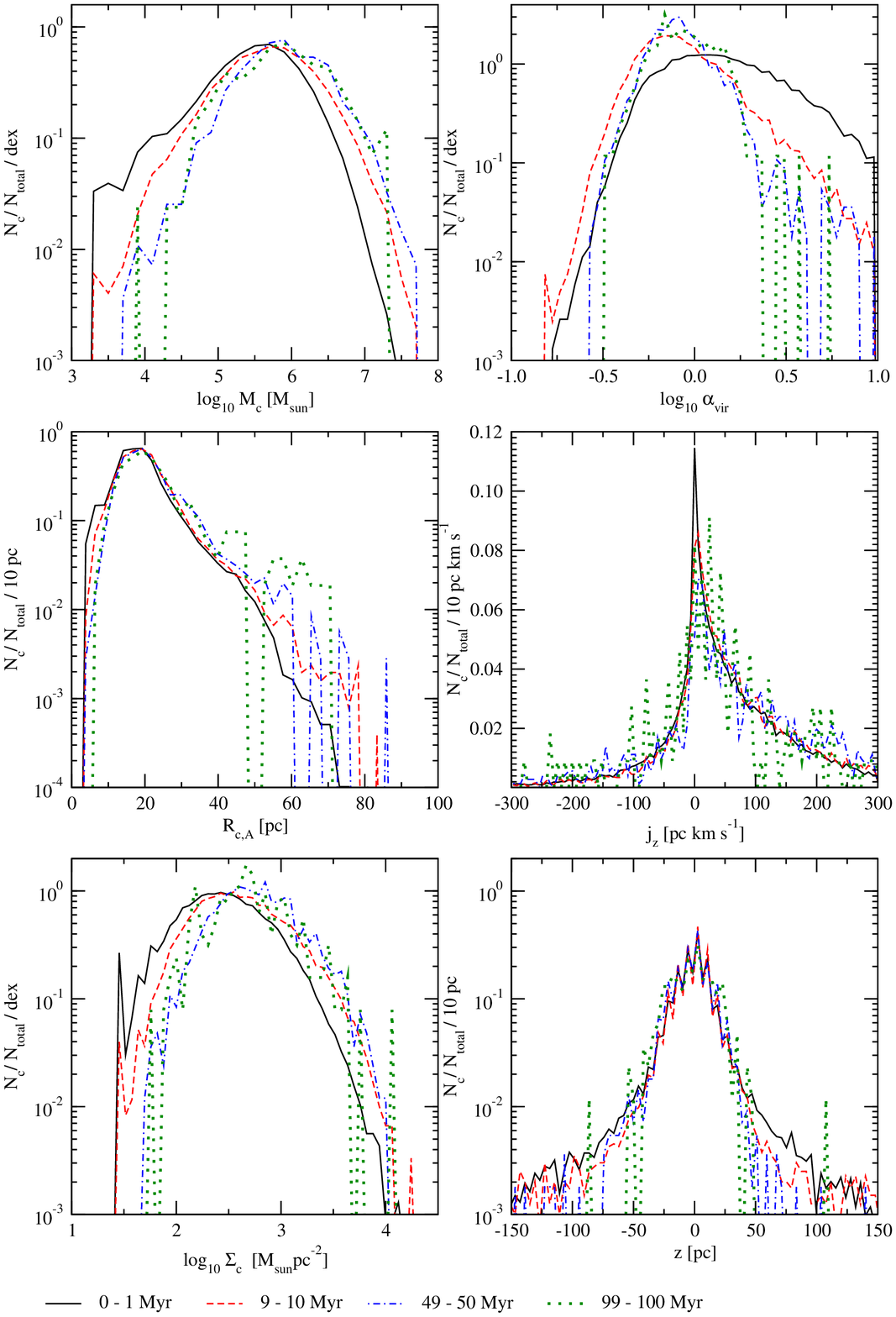}
\caption{
Normalized distributions of GMC properties, as described in
Figure~\ref{fig:simtime_properties}, but now showing results for
different cloud ages: 0-1~Myr (solid lines), 9-10~Myr (dashed lines),
49-50~Myr (dot-dashed lines), 99-100~Myr (dotted lines), for those
clouds born after $t=140$~Myr, i.e. in the fully-fragmented phase.
Top left: (a) Cloud mass, $M_c$. Middle left: (b) Cloud radius,
$R_{c,A} \equiv (A_c/\pi)^{1/2}$. Bottom left: (c) Mass surface
density, $\Sigma_c=M_c/A_c$. Top right: (d) Virial parameter,
$\alpha_{\rm vir}$. Middle right: (e) Vertical ($z$) component of
specific angular momentum, $j_z$.  Bottom right: (f) Cloud center of
mass vertical positions, $z$. 
\label{fig:cloudtime_properties}}
\end{center} 
\end{figure*}

\begin{figure*} 
\begin{center} 
\includegraphics[angle=270,width=\textwidth]{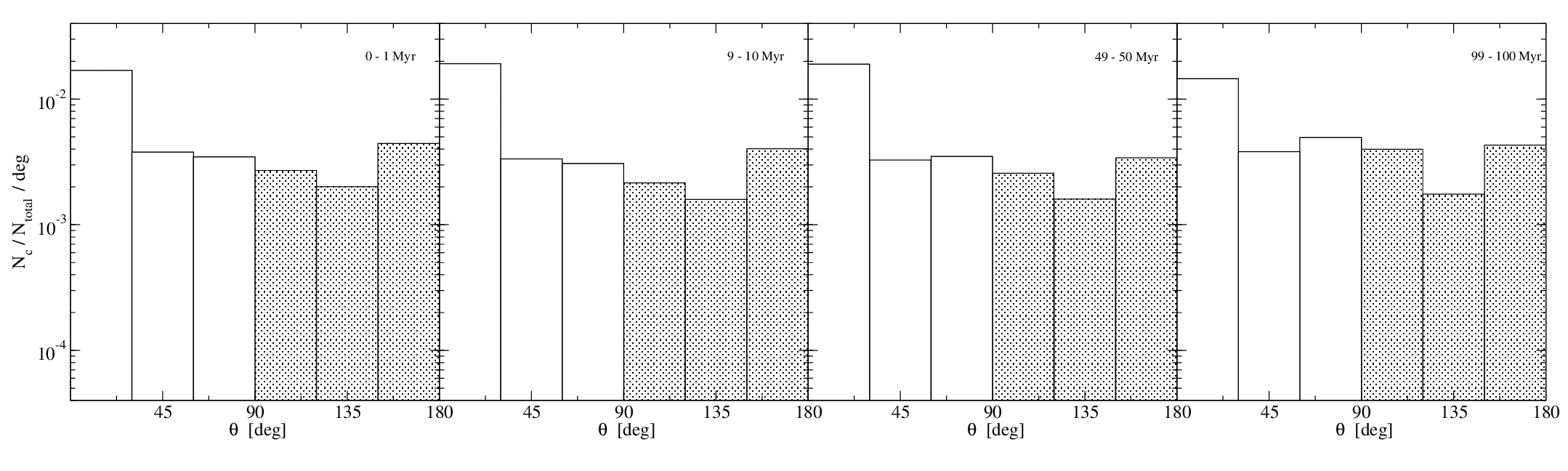}
\caption{Distribution of the angle between cloud angular momentum
vectors and the galactic rotation axis for different cloud ages for
clouds born after $t=140$~Myr in the fully-fragmented stage. The
shaded bars indicated retrograde rotation. The fraction of retrograde
clouds is about 25\% of the total and nearly independent of cloud age.
\label{fig:theta_cloudtime}}
\end{center} 
\end{figure*}

Our method of tracking clouds from one output time to the next allows
us to study the evolution of cloud properties as a function of cloud
age, $t^\prime$. Figure~\ref{fig:cloudtime_properties}, analogous to
Figure~\ref{fig:simtime_properties}, shows how the distribution of
cloud masses, radii, mass surface densities, virial parameters,
vertical component of specific angular momentum, and vertical
positions changes with cloud age ($t^\prime = 0-1, 9-10, 49-50,
99-100$~Myr) for clouds born after $t=140$~Myr, i.e. in the
fully-fragmented phase of the simulation. Younger, especially
newly-formed, clouds tend to have lower masses, slightly smaller
radii, lower mass surface densities, and larger virial parameters
(i.e. are less gravitationally bound). However, young clouds have very
similar distributions of the vertical component of specific angular
momentum and of vertical position above and below the disk midplane.

Figure~\ref{fig:theta_cloudtime} shows the distribution of the angles,
$\theta$, between cloud angular momentum vectors and the galactic
rotation axis for different cloud ages for clouds born after
$t=140$~Myr in the fully-fragmented stage. The fraction of retrograde
clouds is about 25\% of the total and nearly independent of cloud age,
though it has risen to about 30\% for 100~Myr old clouds. At times
$t>140$~Myr, clouds are being born from overdense gas that has already
been disturbed by cloud interactions. At earlier times ($t<140$~Myr)
we find much smaller retrograde fractions, $\sim 10\%$.  These
results, especially for the later simulation times, may depend on the
absence of feedback in the simulation, which would lead to a higher
mass fraction in lower-density gas. Clouds forming from such gas are
expected to form with a higher degree of prograde rotation, since the
initial conditions of the clouds are more dispersed and thus
influenced by galactic shear. The influence of feedback on these
aspects of cloud formation will be investigated in a future paper.

\newpage
\section{Conclusions}

We have presented simulations of GMC formation and evolution in a
marginally unstable gas disk of a flat rotation curve galaxy,
capturing scales from $\sim 20$~kpc down to $\lesssim 10$~pc, with a
fully multiphase atomic ISM. In this initial study, we have focused
on the limit of a negligible star formation rate from the gas and
negligible feedback from those stars. We imposed a minimum effective
sound speed of 1.8~$\rm km\:s^{-1}$ to mimic nonthermal forms of
pressure support in the dense gas. We did not explicitly track
molecule formation on dust grains, but rather defined GMCs to be those
structures with $n_{\rm H}\geq 100\:{\rm cm^{-3}}$, similar to the
mean volume-averaged densities of observed GMCs. Using adaptive mesh
refinement we resolved fragmentation of gas up to densities of about
this cloud threshold density. We developed methods to track clouds
through the simulation, including mergers.

In spite of the simplicity of the model, a surprisingly large number
of observed GMC properties are approximately reproduced by the
simulated cloud population at late simulation times, $t>140$~Myr, once
the disk is fully fragmented. These include the distributions of cloud
mass, size, mass surface density, virial parameter, angular momentum,
vertical height in the disk, line-width size relation and distribution
of angles of angular momentum vectors with respect to the galactic
rotation axis. Many of the ISM and cloud population properties
approach a quasi steady state at late simulation times, indicating an
approximate balance between gravitational instability, heating of the
cloud velocity dispersion via gravitational scattering (and perhaps
some numerical diffusion) and dissipation via cloud collisions and
mergers. The latter processes occur very frequently: a cloud typically
suffers a merger every $1/5$ of an orbital time, i.e. about every
25~Myr at $r=4$~kpc. GMC mergers and collisions, which can be viewed
as being somewhat equivalent to turbulent converging flows in a
self-gravitating and shearing gas disk, are thus efficient at
injecting kinetic energy into the clouds (extracted from orbital
energy) and maintaining near virial balance of the clouds. This
mechanism may be as important as, or even more important than, stellar
feedback, such as momentum injection by supernovae, stellar winds or
ionization, at keeping GMCs turbulent.  A comparison of these
processes will be studied in a future paper.

Cloud collisions may also be efficient triggers for star formation,
since they cause dense gas that is already somewhat prone to star
formation to be pressurized by the ram pressure of the converging
clouds. This could push magnetically subcritical clumps inside GMCs to
become supercritical, thus triggering star cluster formation. Our
result that the cloud collision time is a small and approximately
constant fraction of the local orbital time, lends support to theory of
cloud collision induced star formation to explain the global Kennicutt
(1998) star formation relations, as proposed by Tan (2000). The
correlation of star formation activity with cloud collisions will be
investigated in Paper II.

The complexity of the star formation process means that theoretical
models and numerical simulations must be closely tested against
observed systems. An important future goal is to test how GMC
properties and star formation activity depend on global galactic
properties such as rotation curve shape and velocity normalization,
strength of spiral arm potential, and gas metallicity. Our results
presented here provide a foundation for these future studies.

\acknowledgements We thank P. Barnes, G. Bryan, C. Gammie, N. Gnedin,
A. Kravtsov, M. Krumholz, C. McKee, T. Mouschovias, E. Ostriker,
K. Wada, and T. Wong for helpful discussions. We thank the referee for
their comments and suggestions. EJT acknowledges support from the
Dept. of Astronomy and CLAS, University of Florida via a Postdoctoral
Fellowship in Theoretical Astrophysics. JCT acknowledges support from
NSF CAREER grant AST-0645412. The authors acknowledge the University
of Florida High-Performance Computing Center for providing
computational resources and support that have contributed to the
research results reported within this paper.




\begin{thebibliography}{}

\bibitem[Bergin et al.(2004)]{Bergin2004} Bergin, E.~A., Hartmann, L.~W., Raymond, J.~C., \& Ballesteros-Paredes, J.\ 2004, \apj, 612, 921 

\bibitem[]{bertoldi1992}
Bertoldi, F., \& McKee, C. F. 1992, \apj, 395, 140

\bibitem[]{bigiel08}
Bigiel, F., Leroy, A., Walter, F., Brinks, E., De Blok, W. J. G., Madore, B., \& Thornley, M. D. 2008, \aj, 136, 2846


\bibitem[Binney \& Merrifield(1998)]{BM1998} Binney, J., \& Merrifield, M.\ 1998, Galactic astronomy, Princeton, NJ : Princeton University Press, 1998 

\bibitem[Binney \& Tremaine(1987)]{BT1987} Binney, J., \& Tremaine, S.\ 1987, Princeton, NJ, Princeton University Press, 1987, 747 p.,  

\bibitem[]{blitz90}
Blitz, L. 1990, {\it The Evolution of the Interstellar Medium}, ed. Blitz L., ASP Press: San Francisco, 273


\bibitem[]{blitz2006}
Blitz, L., \& Rosolowsky, E. 2006, \apj, 650, 933

\bibitem[]{bolatto2008}
Bolatto, A., D., Leroy, A. K., Rosolowsky, E., Walter, F., Blitz, L. 2008, \apj, 686, 948

\bibitem[Boulares \& Cox(1990)]{Boulares1990} Boulares, A., \& Cox, D.~P.\ 1990, \apj, 365, 544 


\bibitem[]{bronfmann2000}
Bromfman, L., Casassus, S., May, J., \& Nyman, L. A. 2000, \aap, 358, 521


\bibitem[Bryan \& Norman(1997)]{Bryan1997} Bryan, G.~L.~\& Norman, M.~L.\ 1997, ASP Conf.~Ser.~123: Computational Astrophysics; 12th Kingston Meeting on Theoretical Astrophysics, 363

\bibitem[Bryan(1999)]{Bryan1999}Bryan, G. L. Comp. Phys. and Eng. 1999, 1:2, p.


\bibitem[Cox(2005)]{Cox2005} Cox, D.~P.\ 2005, \araa, 43, 337 

\bibitem[]{crutcher}
Crutcher, R. M. 2005, in {\it Massive star birth: A crossroads of Astrophysics}, IAU Symp. 227, ed. by Cesaroni, R., Felli, M., Churchwell, E., Walmsley, M., (Cambridge: CUP), pp.98

\bibitem[]{dame2001}
Dame, T. M., Hartmann, D., \& Thaddeus, P. 2001, \apj, 547, 792

\bibitem[]{dobbs2008b}
Dobbs, C. L. 2008, \mnras, 391, 844

\bibitem[Dobbs \& Price(2008)]{Dobbs2008} Dobbs, C.~L., \& Price, D.~J.\ 2008, \mnras, 383, 497 



\bibitem[]{elmegreen1994}
Elmegreen, B. G. 1994, \apj, 425, L73

\bibitem[]{elmegreen1986}
Elmegreen, D. M., \& Elmegreen, B. G. 1986, \apj, 311, 554

\bibitem[]{fukui2008}
Fukui, Y., Kawamura, A., Minamidani, T., Mizuno, Y., Kanai, Y. et al. 2008, \apjs, 178, 56

\bibitem[]{gammie1991}
Gammie, C. F., Ostriker, J. P., \& Jog, C. J. 1991, \apj, 378, 565

\bibitem[Glover \& Mac Low(2007a)]{Glover2007a} Glover, S.~C.~O., \& Mac Low, M.-M.\ 2007a, \apjs, 169, 239

\bibitem[Glover \& Mac Low(2007b)]{Glover2007b} Glover, S.~C.~O., \& Mac Low, M.-M.\ 2007b, \apj, 659, 1317

\bibitem[]{gnedin2008}
Gnedin, N. Y., Tassis, K., \& Kravtsov, A. V. 2008, \apj, submitted, (arXiv:0810.4148)

\bibitem[]{grosbol1998}
Grosbol, P. J., \& Patsis, P. A. 1998, \aap, 336, 840

\bibitem[Harfst et al.(2006)]{Harfst2006} Harfst, S., Theis, C., \& Hensler, G.\ 2006, \aap, 449, 509 

\bibitem[Heitsch et al.(2008)]{Heitsch2008} Heitsch, F., Hartmann, L.~W., Slyz, A.~D., Devriendt, J.~E.~G., \& Burkert, A.\ 2008, \apj, 674, 316 

\bibitem[Heitsch et al.(2009)]{Heitsch2009} Heitsch, F., Stone, J., Hartmann, L.~W.\ 2009, \apj, in press (arXiv:0812.3339)

\bibitem[]{heyer2004}
Heyer, M. H., \& Brunt, C. M. 2004, \apj, 615, L45

\bibitem[]{heyer2008}
Heyer, M. H., Krawczyk, C., Duval, J., Jackson, J. M. 2008, \apj, submitted (arXiv:0809.1397)

\bibitem[]{jackson2006}
Jackson, J. M., Rathborne, J. M., Shah, R. Y., Simon, R., Bania, T. M. et al. 2006, \apjs, 163, 145

\bibitem[]{jog}
Jog, C. J. 1996, \mnras, 278, 209

\bibitem[]{jog_sol}
Jog, C. J. \& Solomon, P. M. 1984, \apj, 276, 114



\bibitem[]{joung}
Joung, M. K. R., \& Mac Low, M-M. 2006, \apj, 653, 1266

\bibitem[]{kennicutt1989}
Kennicutt, R. C., Jr. 1989, \apj, 344, 685

\bibitem[Kennicutt(1998)]{Kennicutt1998} Kennicutt, R.~C., Jr.\ 1998, \apj, 498, 541 

\bibitem[]{kennicutt2007}
Kennicutt, R. C., Jr., Calzetti, D., Walter, F., Helou, G., Hollenbach, D. J. et al. 2007, \apj, 671, 333

\bibitem[]{Kim2008} 
Kim, C.-G., Kim, W.-T., \& Ostriker, E.~C.\ 2008, \apj, 681, 1148

\bibitem[Kim \& Ostriker(2001)]{Kim2001} 
Kim, W.-T., \& Ostriker, E.~C.\ 2001, \apj, 559, 70

\bibitem[Kim \& Ostriker(2006)]{Kim2006} 
Kim, W.-T., \& Ostriker, E.~C.\ 2006, \apj, 646, 213 

\bibitem[Kim \& Ostriker(2007)]{Kim2007} 
Kim, W.-T., \& Ostriker, E.~C.\ 2007, \apj, 660, 1232 

\bibitem[Kim et al.(2003)]{Kim2003} 
Kim, W.-T., Ostriker, E.~C., \& Stone, J.~M.\ 2003, \apj, 599, 1157 

\bibitem[]{krumholz2005}
Krumholz, M.~R., \& McKee, C. F. 2005, \apj, 630, 250

\bibitem[Krumholz \& Tan(2007)]{Krumholz2006} 
Krumholz, M.~R., \& Tan, J.~C.\ 2007, \apj, 654, 304

\bibitem[Kwan(1979)]{Kwan1979} Kwan, J.\ 1979, \apj, 229, 567 

\bibitem[]{lada2003}
Lada, C. J., \& Lada, E. A. 2003, \araa, 41, 57

\bibitem[]{larson1988}
Larson, R.B. 1988, in {\it Galactic and Extragalactic Star Formation}, ed. R.E. Pudritz \& M. Fich, Dordrecht: Kluwer, 435

\bibitem[]{leroy08}
Leroy, A. K., Walter, F., Brinks, E., Bigiel, F., de Blok, W. J. G., Madore, B., \& Thornley, M. D. 2008, \aj, 136, 2782

\bibitem[Li et al.(2005)]{Li2005} Li, Y., Mac Low, M.-M., \& Klessen, R.~S.\ 2005, \apjl, 620, L19 

\bibitem[Li et al.(2006)]{Li2006} 
Li, Y., Mac Low, M.-M., \& Klessen, R.~S.\ 2006, \apj, 639, 879 

\bibitem[]{Martin2001}
Martin, C. L., \& Kennicutt, R. C. 2001, \apj, 555, 301

\bibitem[]{Matzner2002}
Matzner, C. D. 2002, \apj, 566, 302

\bibitem[]{mccall1986}
McCall, M. L., \& Schmidt, F. H. 1986, \apj, 311, 548

\bibitem[McKee \& Ostriker(2007)]{McKee2007} McKee, C.~F., \& Ostriker, E.~C.\ 2007, \araa, 45, 565

\bibitem[McKee \& Ostriker(1977)]{McKee1977} McKee, C.~F., \& Ostriker, J.~P.\ 1977, \apj, 218, 148

\bibitem[]{mckee1997}
McKee. C. F., \& Williams, J. P. 1997, \apj, 476, 144

\bibitem[O'Shea et al.(2004)]{OShea2004} O'Shea, B.~W., Bryan, G., Bordner, J., Norman, M.~L., Abel, T., Harkness, R., \& Kritsuk, A.\ 2004, ArXiv Astrophysics e-prints, astro-ph/0403044

\bibitem[]{padoan2002}
Padoan, P., \& Nordlund, A. 2002, \apj, 576, 870

\bibitem[]{padoan1997}
Padoan, P., Nordlund, A., \& Jones, B. J. T. 1997, \mnras, 288, 145

\bibitem[]{phillips1999}
Phillips, J. P. 1999, \aaps, 134, 241


\bibitem[]{robertson2008}
Robertson, B. E., \& Kravtsov, A. V. 2008, \apj, 680, 1083

\bibitem[Robertson et al.(2004)]{Robertson2004} Robertson, B., Yoshida, N., Springel, V., \& Hernquist, L.\ 2004, \apj, 606, 32



\bibitem[Rosen \& Bregman(1995)]{Rosen1995} Rosen, A.~\& Bregman, J.~N.\ 1995, \apj, 440, 634 

\bibitem[Rosolowsky et al.(2003)]{Rosolowsky2003} Rosolowsky, E., Engargiola, G., Plambeck, R., \& Blitz, L.\ 2003, \apj, 599, 258 

\bibitem[Sarazin \& White(1987)]{Sarazin1987} Sarazin, C.~L.~\& White, R.~E.\ 1987, \apj, 320, 32 

\bibitem[]{schaye2004}
Schaye, J. 2004, \apj, 609, 667

\bibitem[]{seiden1990}
Seiden, P. E., \& Schulman, L. S. 1990, Adv. Phys., 39, 1

\bibitem[Shetty \& Ostriker(2008)]{Shetty2008} Shetty, R., \& Ostriker, E.~C.\ 2008, \apj, 684, 978 

\bibitem[Shetty \& Ostriker(2006)]{Shetty2006} Shetty, R., \& Ostriker, E.~C.\ 2006, \apj, 647, 997 

\bibitem[Slyz et al.(2005)]{Slyz2005} Slyz, A.~D., Devriendt, J.~E.~G., Bryan, G., \& Silk, J.\ 2005, \mnras, 356, 737 

\bibitem[]{solomon1987}
Solomon, P. M., Rivolo, A. R., Barrett, J., Yahil, A. 1987, \apj, 319, 730

\bibitem[]{stark1989}
Stark, A. A., \& Brand, J. 1989, \apj, 339, 763

\bibitem[]{stark2005}
Stark, A. A, \& Lee, Y. 2005, \apj, 619, L159

\bibitem[Stone \& Norman(1992)]{Stone1992} Stone, J.~M.~\& Norman, M.~L.\ 1992, \apjs, 80, 753 

\bibitem[Tan(2000)]{Tan2000} Tan, J.~C.\ 2000, \apj, 536, 173

\bibitem[Tasker \& Bryan(2006)]{Tasker2006} Tasker, E.~J., \& Bryan, G.~L.\ 2006, \apj, 641, 878

\bibitem[Tasker \& Bryan(2008)]{Tasker2008} Tasker, E.~J., \& Bryan, G.~L.\ 2008, \apj, 673, 810 

\bibitem[Tasker et al.(2008)]{Tasker2008b} Tasker, E.~J., Brunino, R., Mitchell, N.~L., Michielsen, D., Hopton, S., Pearce, F.~R., Bryan, G.~L., \& Theuns, T.\ 2008, \mnras, submitted

\bibitem[]{thornley1997a}
Thornley, M. D., \& Mundy, L. G. 1997a, \apj, 484, 202

\bibitem[]{thornley1997b}
Thornley, M. D., \& Mundy, L. G. 1997b, \apj, 490, 682

\bibitem[Toomre(1964)]{Toomre1964} Toomre, A.\ 1964, \apj, 139, 1217 

\bibitem[Truelove et al.(1997)]{Truelove1997} Truelove, J.~K., Klein, R.~I., McKee, C.~F., Holliman, J.~H., Howell, L.~H., \& Greenough, J.~A.\ 1997, \apjl, 489, L179 

\bibitem[Wada \& Norman(2001)]{Wada2001} Wada, K., \& Norman, C.~A.\ 2001, \apj, 547, 172 

\bibitem[Wada \& Norman(2007)]{Wada2007} Wada, K., \& Norman, C.~A.\ 2007, \apj, 660, 276

\bibitem[]{wang1994}
Wang, B., \& Silk, J. 1994, \apj, 427, 759

\bibitem[Wannier et al.(1991)]{Wannier1991} Wannier, P.~G., Lichten, S.~M., Andersson, B.-G., \& Morris, M.\ 1991, \apjs, 75, 987 

\bibitem[Williams \& McKee(1997)]{Williams1997} Williams, J.~P., \& McKee, C.~F.\ 1997, \apj, 476, 166 

\bibitem[Wolfire et al.(2003)]{Wolfire2003} Wolfire, M.~G., McKee, C.~F., Hollenbach, D., \& Tielens, A.~G.~G.~M.\ 2003, \apj, 587, 278

\bibitem[]{wong2002}
Wong, T., \& Blitz, L. 2002, \apj, 569, 157

\bibitem[]{wyse1986}
Wyse, R. F. G. 1986, \apj, 311, L41

\bibitem[]{wyse1989}
Wyse, R. F. G., \& Silk, J. 1989, \apj, 339, 700

\bibitem[]{zuckerman1974}
Zuckerman, B., \& Evans, N. J., II. 1974, \apj, 192, L149

\end{thebibliography}
\end{document}